\documentclass{vgtc}                          

\graphicspath{{figures/}{pictures/}{images/}{./}} 

\usepackage{times}                     

\usepackage{tabu}                      
\usepackage{booktabs}                  
\usepackage{lipsum}                    
\usepackage{mwe}                       

\usepackage{mathptmx}                  

\usepackage{tabularx}
\usepackage{multirow}
\usepackage{cite}
\usepackage{amsmath,amssymb,amsfonts}
\usepackage{algorithmic}
\usepackage{graphicx}
\usepackage{float}
\usepackage{textcomp}
\usepackage{xspace}
\usepackage[caption=false]{subfig}
\usepackage{enumitem}

\newcommand{\sysname}{$\sf{\it EVER}$\xspace}

\onlineid{1968}
\vgtccategory{ISMAR 2025 Papers}
\vgtcinsertpkg

\title{EVER: Edge-Assisted Auto-Verification for Mobile MR-Aided Operation}

\author{Jiangong Chen\thanks{co-first author. e-mail: jiangong@psu.edu} %
\and Mingyu Zhu\thanks{co-first author. e-mail: mintrrey@psu.edu} 
\and Bin Li\thanks{e-mail: binli@psu.edu}
} %

\affiliation{\scriptsize Department of Electrical Engineering, Pennsylvania State University, University Park}

\teaser{
  \centering
  \includegraphics[width=\textwidth]{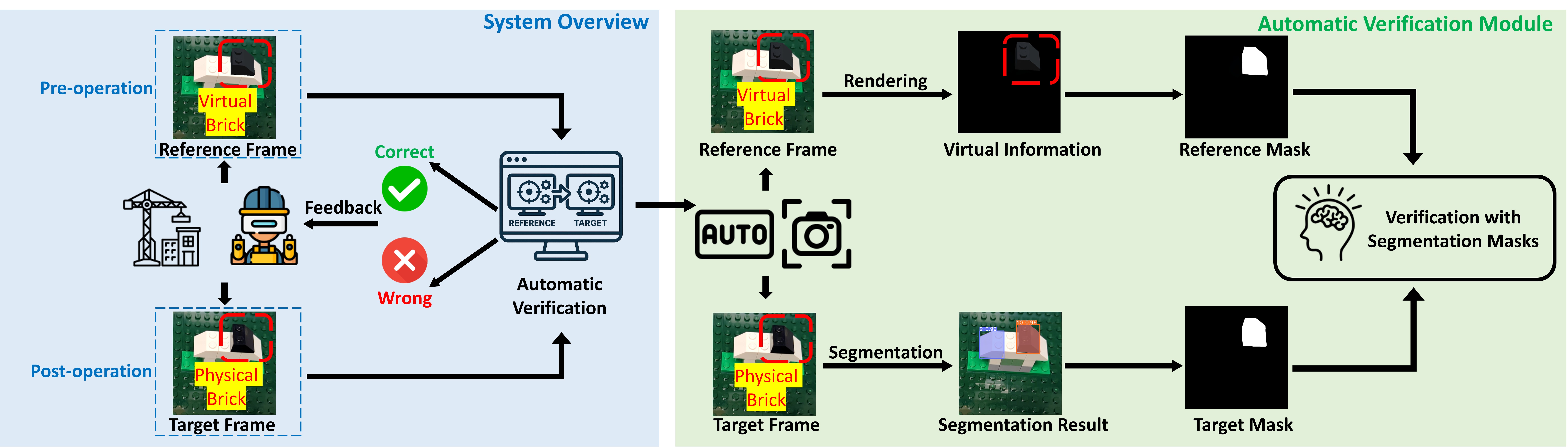}
  \caption{An illustrative example of auto-verification in an MR-aided operation system. The headset automatically captures the frame before and after the user's action, leverages deep learning models and virtual information to convert frames into segmentation masks, and displays immediate feedback on whether the user takes the correct action.
  }
  \label{fig:teaser}
}

\abstract{
Mixed Reality (MR)-aided operation overlays digital objects on the physical world to provide a more immersive and intuitive operation process. A primary challenge is the precise and fast auto-verification of whether the user follows MR guidance by comparing frames before and after each operation. The pre-operation frame includes virtual guiding objects, while the post-operation frame contains physical counterparts. Existing approaches fall short of accounting for the discrepancies between physical and virtual objects due to imperfect 3D modeling or lighting estimation.
In this paper, we propose EVER: an edge-assisted auto-verification system for mobile MR-aided operations. Unlike traditional frame-based similarity comparisons, EVER leverages the segmentation model and rendering pipeline adapted to the unique attributes of frames with physical pieces and those with their virtual counterparts; it adopts a threshold-based strategy using Intersection over Union (IoU) metrics for accurate auto-verification. To ensure fast auto-verification and low energy consumption, EVER offloads compute-intensive tasks to an edge server. Through comprehensive evaluations of public datasets and custom datasets with practical implementation, EVER achieves over 90\% verification accuracy within 100 milliseconds (significantly faster than average human reaction time of approximately 273 milliseconds), while consuming only minimal additional computational resources and energy compared to a system without auto-verification.
} 

\begin{document}

\firstsection{Introduction}

\maketitle

Mixed Reality (MR)-aided operation, which fuses the digital elements with physical world settings, holds great potential to revolutionize how tasks are approached in various scenarios, including laboratory operations \cite{smith2016augmented}, manufacturing processes \cite{sahija2021impact}, maintenance activities \cite{naticchia2018mixed}, and remote guidance systems \cite{thoravi2019loki}.
MR-aided operation system allows users to interact with a mixed environment, where virtual guidance and panels are superimposed onto physical ones, indicating the precise placement of parts to be assembled. This provides an efficient and intuitive alternative to conventional in-person instruction or assembly manuals, significantly enhancing operators' productivity and reliability. For instance, aerospace giant Boeing has experimented with Google Glass Enterprise Edition to streamline its assembly process, achieving a 25\% reduction in assembly time and a notable drop in error rates (see \cite{greengard2019virtual}). 

A crucial aspect of an effective MR-aided operation system lies in its ability to automatically validate the user's adherence to MR directives.
This involves comparing pre-operation frames, which include virtual guidance, with post-operation frames, which display the physical counterpart of the MR directives. 
The auto-verification process offers timely feedback on user actions, reducing the need for manual inspection and minimizing the risk of cumulative errors. However, integrating auto-verification into an MR-aided operation system presents several challenges:
i) Imperfect target object modeling and lighting estimation make traditional similarity-based analysis unsuitable for accurately comparing virtual objects with their physical counterparts.
ii) The dynamic nature of user behavior while wearing MR headsets makes it difficult to capture accurate frames for comparison, particularly in the presence of hand occlusions and head movements.
iii) Machine learning (ML)-based object detection and segmentation models often involve high computational costs and latency, leading to a degraded user experience.

In light of those challenges, the following research questions arise:
i) How can we leverage the unique properties of MR contexts to validate user actions despite incomplete or imperfect knowledge?
ii) How can we ensure the reliability of verification results across complex environments, diverse application scenarios, and dynamic user behaviors?
iii) How can we provide prompt feedback within the span of human reaction time (approximately 273 ms, see \cite{humanreaction}) to ensure a seamless user experience?
iv) How can we minimize energy consumption on mobile devices to prevent overheating and battery drain?
Although auto-verification is a well-established concept in fields such as language programming and computing systems (e.g., \cite{igarashi1975automatic,clarke1986automatic,alur1991techniques}), its application within MR-aided operation remains largely unexplored. Previous studies on MR-aided operation have primarily focused on the precise localization of virtual models (e.g., \cite{yan2021augmented}) or synchronizing design progress among multiple users (e.g., \cite{yao2022scalable}). However, these approaches rely on manual control or simplistic triggers, such as hand detection, to progress guidance steps without auto-verification. While some works (e.g., \cite{yao2023design}) have considered automatic verification, their approach is grounded in traditional image similarity checks. Such methods fail to adequately address the inherent differences between digital and physical elements in MR, leading to unreliable verification results. 

In this paper, we propose \sysname: an \underline{E}dge-assisted auto-\underline{ver}ification system for mobile MR-aided operation applications. \sysname enhances user experience by accurately displaying virtual objects at designated locations and autonomously verifying the user's adherence to the provided instructions. Unlike traditional methods that primarily rely on similarity comparisons, 
\sysname considers both similarities and discrepancies between digital and physical objects in MR-aided operation applications and presents an approach that leverages rendering information to enhance verification robustness. 
We have designed an end-to-end and fully automated system, chaining together multiple technical layers, including graphics rendering, visual perception, compression, and networking, each carefully tuned to balance latency and accuracy. It seamlessly integrates motion detection, object segmentation, and edge computing into a unified and extensible framework that addresses key challenges in real-time MR-aided operation.
The key design aspects of \sysname are summarized as follows. 

First, we design an automated motion detection method to identify critical `reference' and `target' frames based on the user's behavior in MR-aided operation environments. These frames represent the user's operations before and after MR guidance, respectively. Second, \sysname employs the rendering pipeline and learning-based segmentation model to obtain segmentation masks for digital objects and their physical counterparts.
Moreover, we present a threshold-based policy anchored in the Intersection over Union (IoU) metric (see \cite{van2019deep}) derived from the segmentation outcomes. Third, we integrate edge computing in \sysname and implement a series of optimization techniques to ensure its practicality, reduce energy consumption, and augment system responsiveness. 

Overall, \sysname stands out by offering a fully automated, end-to-end, high-accuracy, low-latency, and energy-efficient solution for verifications in mobile MR-aided operation fields.
We have provided the fully deployable implementation and comprehensive evaluations on over 20,000 frames from both public and custom datasets, achieving a remarkable verification accuracy of approximately 90\%.
Moreover, in practical implementation, our system achieves an end-to-end latency of under 100 milliseconds, much smaller than the average human reaction time (around 273 milliseconds), consuming a minimal amount of computational resources and energy compared to that without auto-verification. The main contributions of our paper are summarized below:

\begin{itemize}
    \item We present \sysname, a unified and extensible framework for end-to-end, fully automated verification in mobile MR-aided operation tasks, detailing its core modules, auto-verification approach, and practical implementation techniques.
    \item We evaluate the proposed auto-verification approach on large-scale synthetic datasets based on public datasets to demonstrate the generality. Those datasets simulate laboratory operations within MR-aided contexts, specifically targeting PCB boards and breadboards.
    \item We also collect a custom LEGO dataset to fine-tune ML models and deploy \sysname on commodity mobile devices. We conduct extensive evaluations and benchmarks to demonstrate its efficiency and responsiveness.
\end{itemize}

The remainder of this paper is organized as follows:
\Cref{sec::related} summarizes existing literature in related fields. \Cref{sec::arch} introduces the system architecture. \Cref{sec::design} presents a detailed description of the design principles and optimization techniques applied. \Cref{sec::imple} provides the software and hardware implementation. \Cref{sec::eva} shows the evaluation results based on various datasets and our system implementation. \Cref{sec::discuss} discusses the limitations and future directions and \Cref{sec::conclusion} concludes the paper.

\section{Related Work} \label{sec::related}
This section reviews three related topics: edge-assisted mobile system, automatic verification, and MR-aided operation. We summarize previous achievements and indicate their approaches' limitations when applied to the scenarios considered in this paper.

\textbf{Edge-assisted Mobile System.}
Offloading intensive computational tasks to an edge server has proven to be a feasible way applied in many fields. For example, real-time object recognition can be realized on mobile devices by offloading the deep-learning tasks to an edge server (see \cite{chen2015glimpse,liu2019edge}). Researchers have also made significant efforts to reduce communication and computation latency in edge computing applications (see \cite{abbas2017mobile,chen2022enhancing,chen2023motion}).
However, auto-verification in MR-aided operation implies a highly complicated system with multiple tightly coupled components and thus presents unique challenges and opportunities for determining the allocation of modules between the client and server sides. In this paper, we propose a novel edge-assisted system design that seamlessly integrates those modules while minimizing computational overhead and energy consumption on mobile devices.

\textbf{Automatic Verification.}
Auto-verification refers to the process of using automated systems to verify the accuracy, integrity, or validity of data, algorithms, or designs. This technique has been widely accepted and explored by researchers (e.g., \cite{igarashi1975automatic,clarke1986automatic,alur1991techniques})
to eliminate the need for manual checks. In the context of MR-aided operations, auto-verification takes on aspects similar to traditional image classification (see \cite{lu2007survey,druzhkov2016survey}) and similarity-checking problems (\cite{brown1992survey,thung2009survey}). However, existing methods predominantly address either purely physical or entirely virtual images. The unique property of auto-verification in MR-aided operations lies in comparing images blending both virtual and physical elements, presenting distinct challenges and opportunities elaborated in this paper.

\textbf{MR-aided Operation.}
Integrating MR technologies in operations significantly enhances user immersion and operational efficiency. However, the high cost of advanced MR headsets has prompted researchers (e.g., \cite{yan2021augmented, yao2022scalable, dong2023collaborative}) to explore precise localization and improved collaboration on more affordable devices. While these studies have improved user experiences, they often overlook the crucial aspect of auto-verification in MR-aided operations, essential for reducing manual intervention and ensuring a seamless, immersive experience.
Some studies (see \cite{yao2023design}) have begun exploring auto-verification within this domain. However, these efforts often fall short of providing comprehensive technical analysis, treating the issue as a mere similarity check problem. This oversimplification results in approaches that lack the necessary robustness and reliability for MR-aided operations.
In the next section, we will introduce the design principles of \sysname and the optimizations to ensure efficiency and responsiveness.

\section{System Architecture} \label{sec::arch}

\begin{figure}[ht]
    \vspace{-0.1in}
    \centering
    \includegraphics[width=0.48\textwidth]{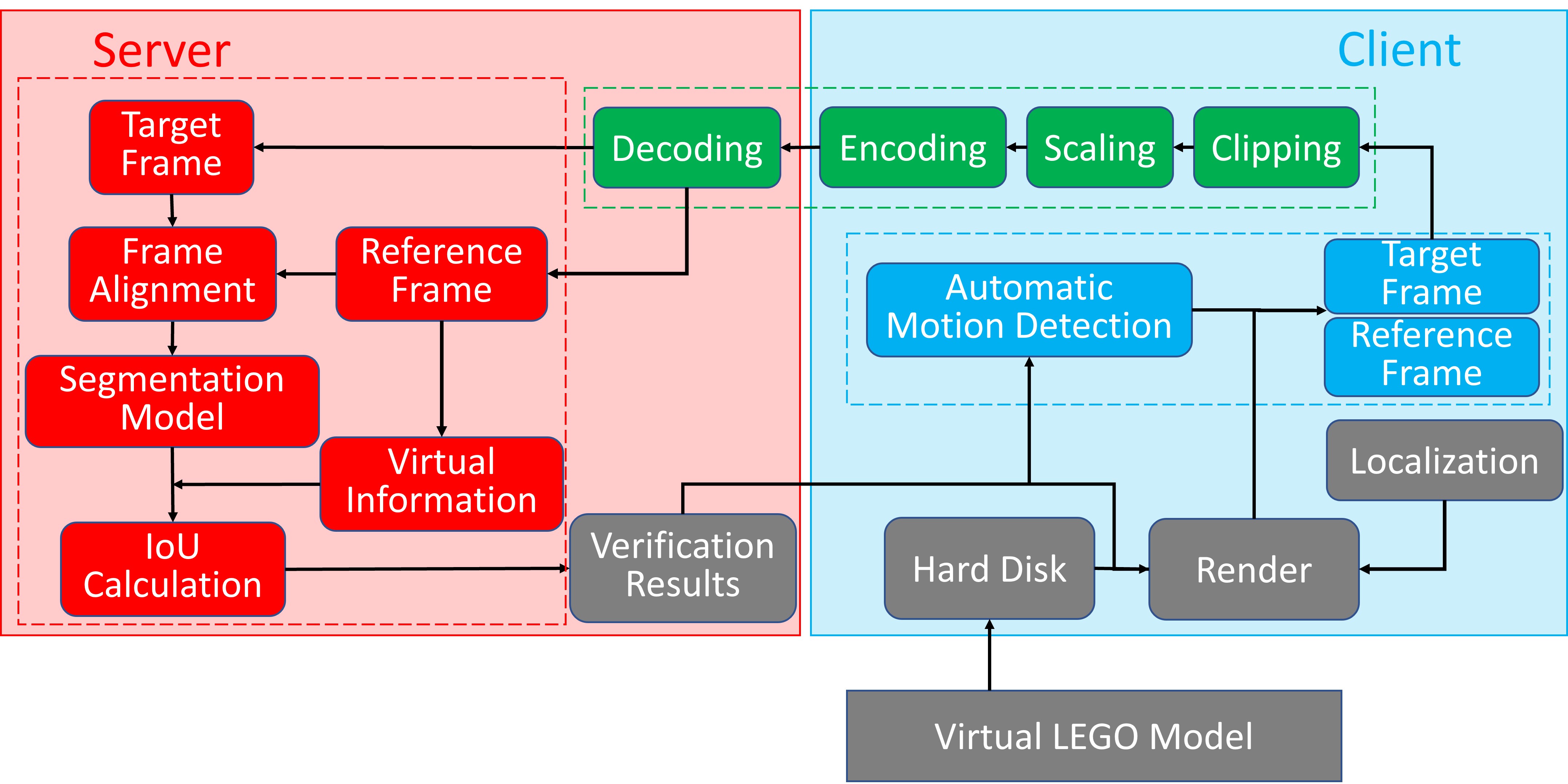}
    \caption{{System architecture.}}
    \label{fig:detection}
    \vspace{-0.15in}
\end{figure}

In this section, we provide an overview of \sysname, which provides MR-aided operation guidance and automatic verification. Although our implementation focuses on the LEGO construction system due to its widespread accessibility, the design principles are applicable to other MR-aided operation systems. \sysname provides guidance to the user on where and which LEGO piece should be localized in each step by displaying a virtual LEGO block over the physical LEGO units. Besides, \sysname supports fast and accurate auto-verification that provides timely feedback on whether the user puts the correct LEGO piece in the desired location. \Cref{fig:detection} illustrates our system architecture consisting of an edge server and a mobile user, which communicate via WiFi. 

On the user's end, our system displays the virtual brick upon existing physical LEGO units to guide the user in placing the corresponding physical piece in the desired location. To achieve accurate localization of virtual bricks, we adopt the open-source framework AprilTag (see \cite{olson2011apriltag,wang2016apriltag}), and use white and black LEGO bricks to build the tag.
We store the whole 3D model on the hard disk while only displaying the next-step piece to enable fast virtual brick rendering. We also deploy a motion detection module to automatically detect whether the user has finished the current step. A reference frame containing the rendered virtual brick and a target frame including only physical bricks are captured before and after the user puts a corresponding physical piece, respectively. Then, both frames are uploaded to the edge server to verify whether the user's placement is correct. We perform image cropping, scaling, and encoding before uploading to the edge server to reduce processing latency while maintaining good verification performance. After receiving the verification result from the edge server, the system moves to the next-step virtual brick if the user correctly places the brick; otherwise, it remains in the current step. 

On the edge server's side, upon receiving the byte stream and performing decoding, it employs different strategies to segment the virtual and physical pieces in the reference and target frames, respectively. Then, it calculates the IoU of the segmented masks to perform the verification based on a threshold-based policy.
Once automatic verification is completed, the edge server sends the outcome to the mobile user. In the next section, we will illustrate the principles and detailed techniques of our system design. 

\section{System Design}
\label{sec::design}

In this section, we showcase the main designs of \sysname. First, we propose the automatic motion detection method to capture the user's operation. Next, we highlight the principles of the auto-verification method. Lastly, we describe some optimizations to ensure \sysname works in practical scenarios and enhance its efficiency and responsiveness.

\subsection{Automatic Motion Detection}
\begin{figure}[ht]
    \vspace{-0.2in}
    \centering
    \subfloat[Reference frame.]{\includegraphics[width=0.15\textwidth]{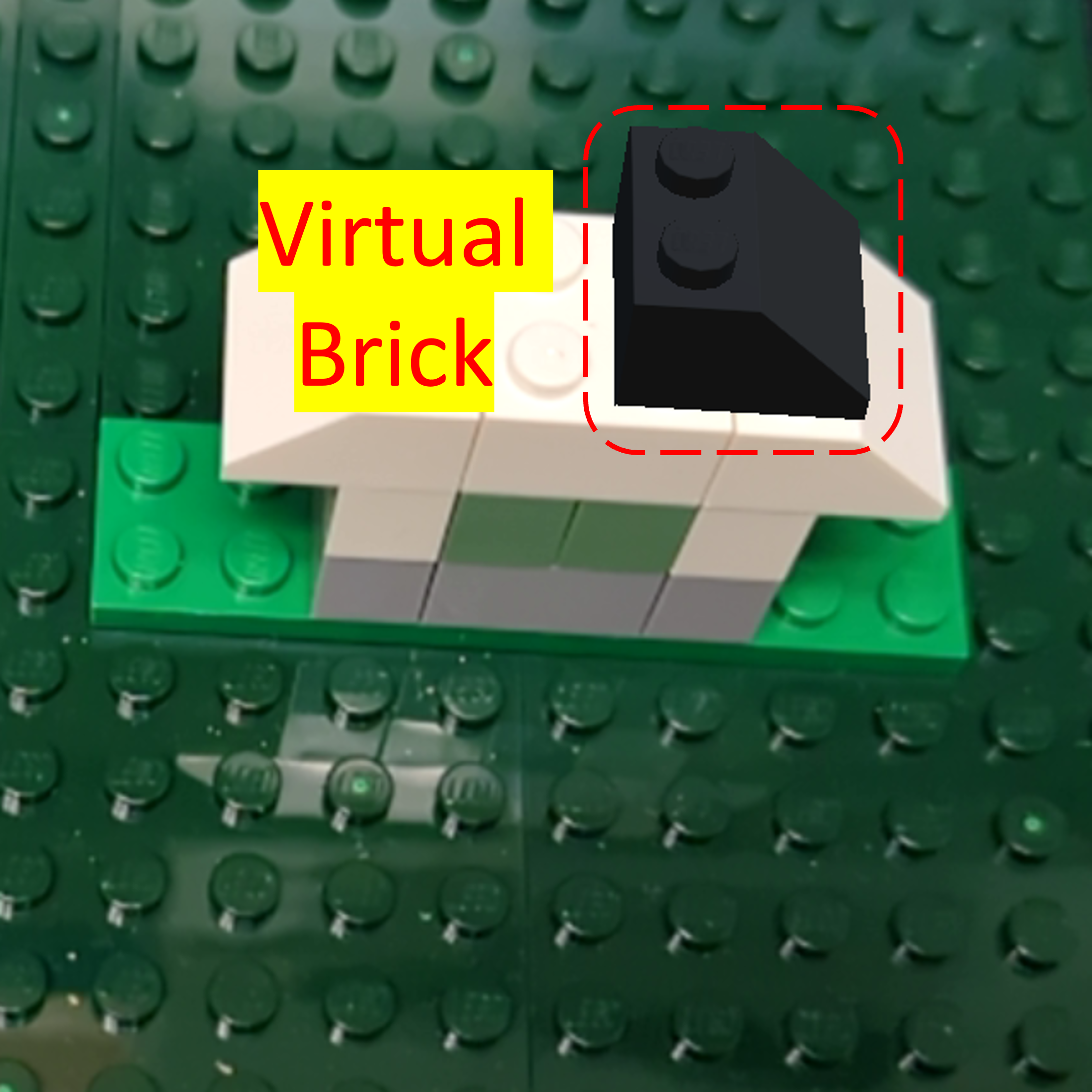}
    \label{fig:ex-ref}}\hfill
    \subfloat[Virtual information.]{\includegraphics[height=0.15\textwidth]{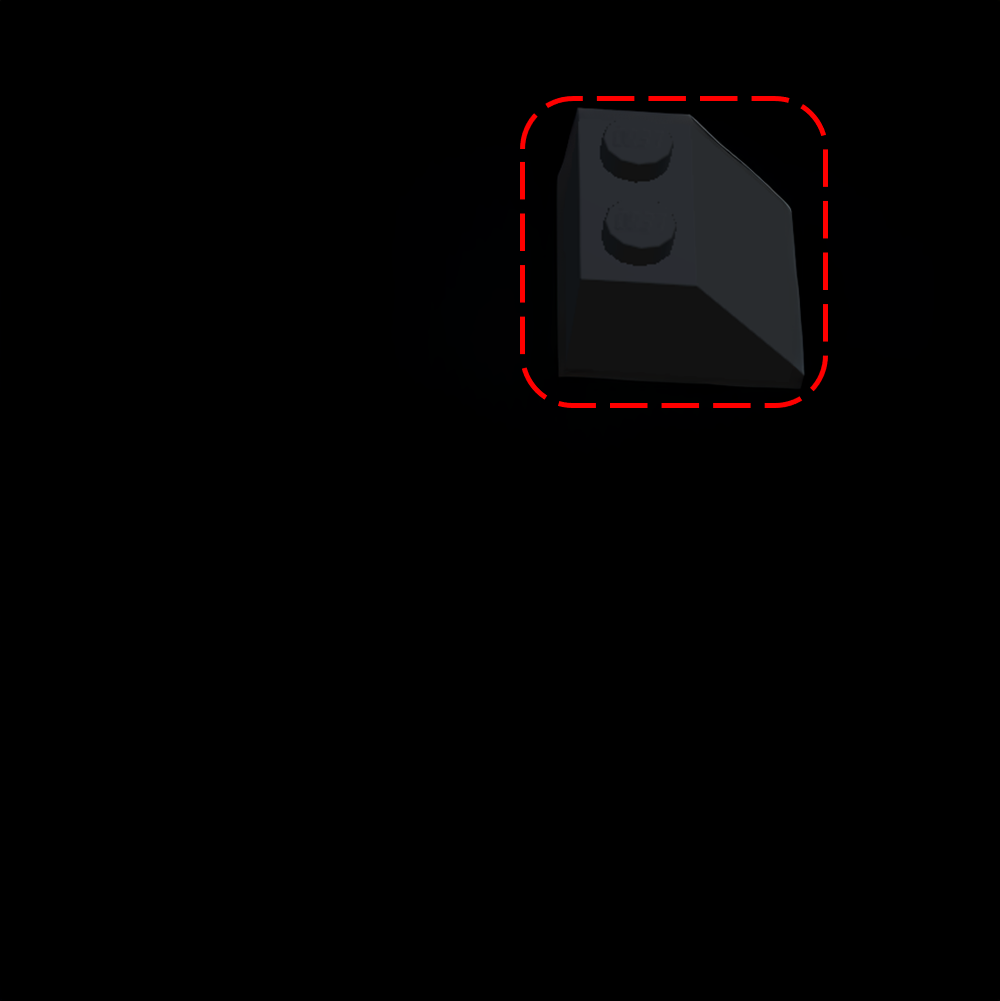}
    \label{fig:virtual_info}}\hfill
    \subfloat[{Mask.}]{\includegraphics[height=0.15\textwidth]{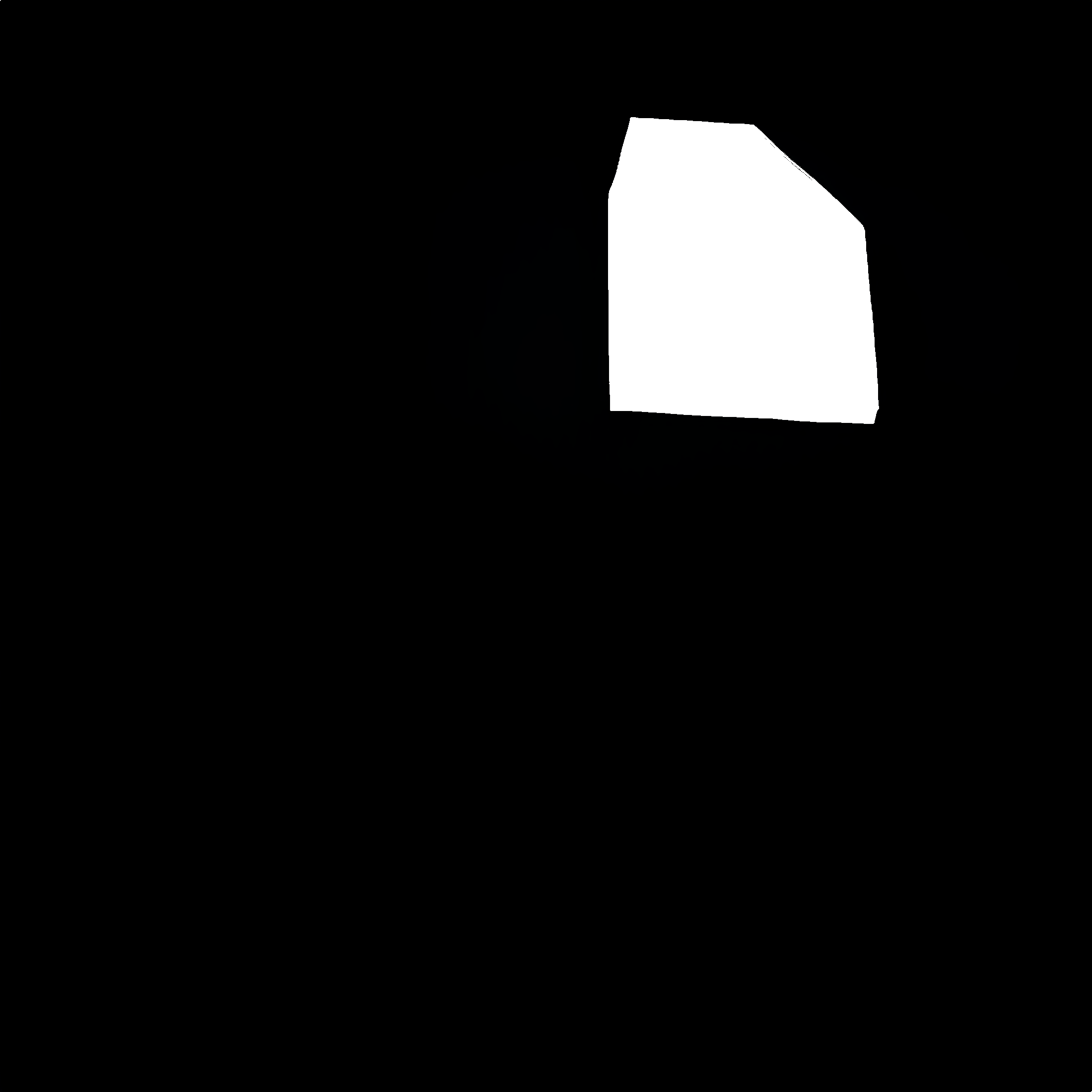}
    \label{fig:ref_mask}} \\
    \subfloat[{Target frame.}]{\includegraphics[width=0.15\textwidth]{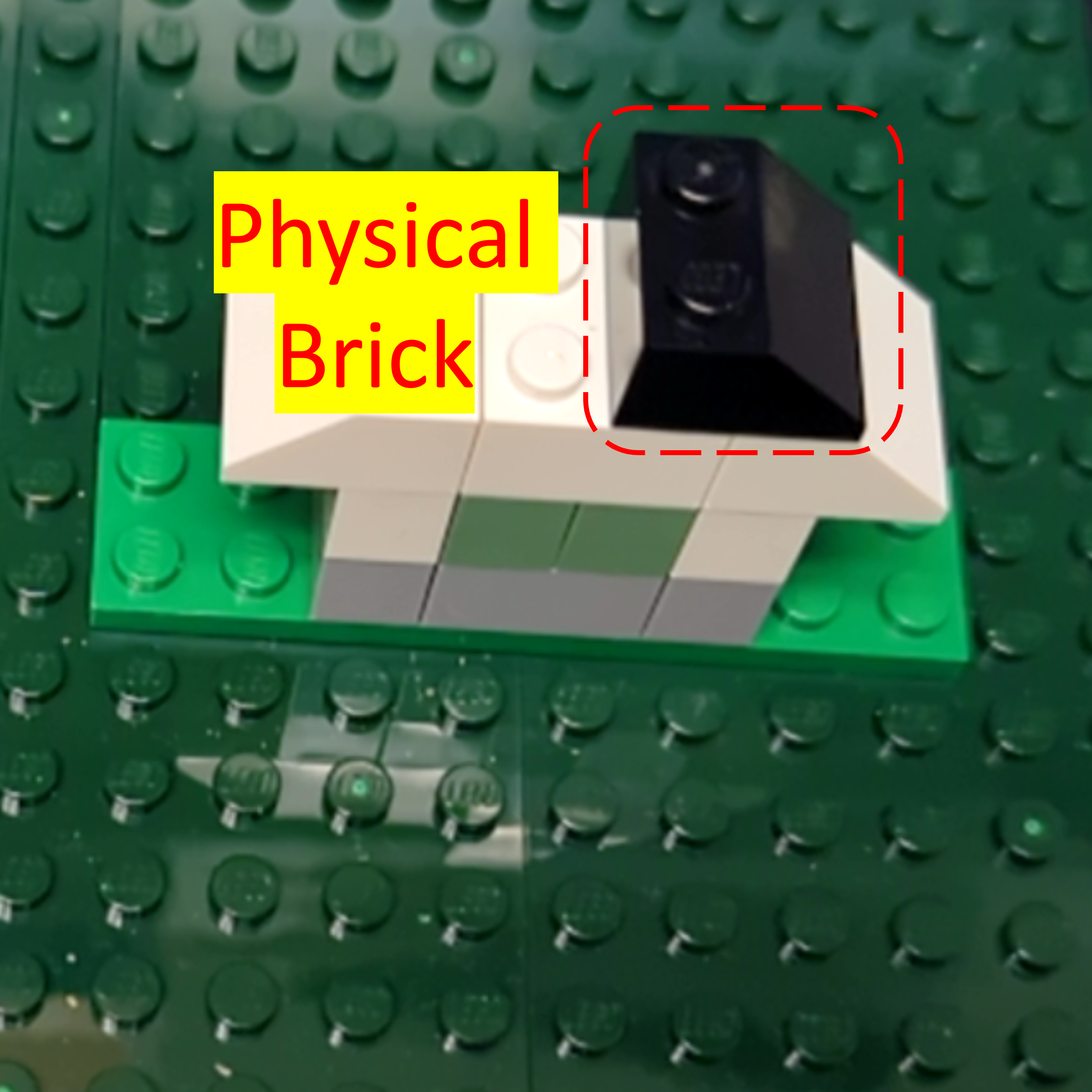}
    \label{fig:ex-tar}}\hfill
    \subfloat[YOLO result.]{\includegraphics[height=0.15\textwidth]{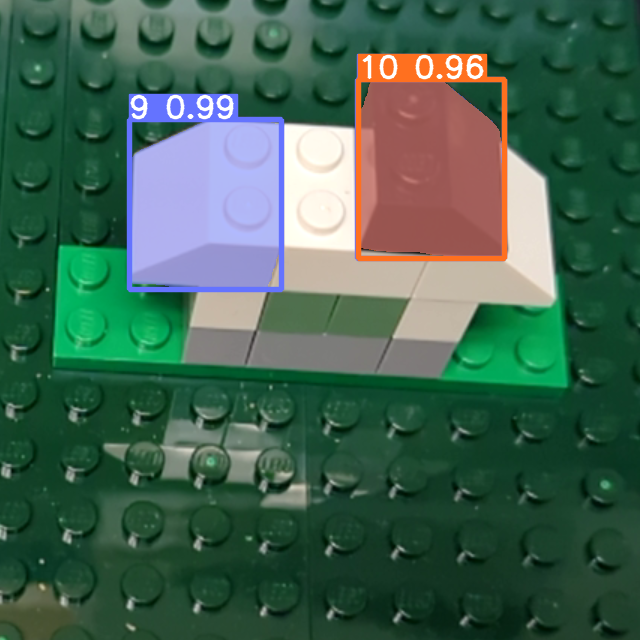}
    \label{fig:seg_results:pmodel}}\hfill
    \subfloat[YOLO mask.]{\includegraphics[height=0.15\textwidth]{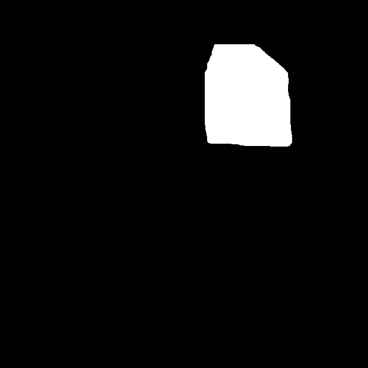}
    \label{fig:seg_masks:pmodel}}
    \caption{Example frames correctly following MR instructions.
    }
    \label{fig:ex-frames}
    \vspace{-0.1in}
\end{figure}

To support the verification process for LEGO construction, we capture two screenshots at each construction step, called the reference frame and the target frame, respectively. The \emph{reference frame} refers to the screenshot captured before the user's operation, which includes the virtual LEGO model that instructs the user's next-step LEGO brick, as shown in \Cref{fig:ex-ref}. 
Note that only the next step of the construction needs to be rendered, so the rendering load remains lightweight regardless of the model size. The \emph{target frame} is the screenshot taken after the user completes her current guided construction step and only contains the desired physical brick, as depicted in \Cref{fig:ex-tar}. Notably, we can observe imperfect color modeling and lightning estimation when comparing those two frames. We will demonstrate that our auto-verification approach only requires reasonable shape modeling and is robust to those imperfect factors in \Cref{sec::autoveri}. 

The reference and target frame pair must be captured at the appropriate time to ensure auto-verification accuracy. To address this, we propose a method to track the user’s actions and prevent frame capture when the user’s body occludes the bricks. Note that the workflow of LEGO construction involves the user's hands entering the camera’s field of view (FoV) to place a brick in the target position and then exiting the frame to retrieve new bricks after completing the step. Based on this observation, we design an automatic motion detection approach using hand detection to determine when the user has completed the current operation.

Specifically, we model the user’s behavior in two stages: idle and busy. Initially, the user is in the idle stage, during which we capture a reference screenshot for subsequent comparisons. We then periodically capture new frames and employ a lightweight hand detector (see \cite{yan2021augmented}, Section 4.5.1) based on skin color detection (see \cite{shaik2015comparative}). When a hand is detected, defined as the proportion of pixels falling into the range of skin color exceeding a threshold, the user transitions to the busy stage. The disappearance of hands during the busy stage signals a return to the idle state, indicating the completion of the construction step and triggering the capture of target frames. After the server processes the target frames and sends feedback to the user, the displayed content is refreshed, and new reference frames are captured to repeat this process.
To enhance robustness, we dynamically adjust the hand-detection threshold based on the distance between the QR code and the camera, as the hand may appear smaller when the LEGO bricks are farther away. Additionally, the responsiveness of the system is influenced by the capture period: a shorter period results in quicker responses but incurs higher computational overhead. Considering human reaction time and operational efficiency, we set the capture period to 100 ms.

\subsection{Automatic Verification} \label{sec::autoveri}
In this section, we propose our design to address key challenges of auto-verification under certain assumptions, such as precise placement of virtual guidance and proper alignment of reference and target frames. These conditions will be addressed and fulfilled in \Cref{sec::practical}, where we adapt the system for more practical scenarios.

A straightforward approach to implementing auto-verification is to compare the reference and target frames using direct frame difference calculations or image similarity metrics, such as Structural Similarity Index (SSIM) and Peak Signal-to-Noise Ratio (PSNR). Intuitively, adherence to MR guidance yields smaller differences, while unexpected behaviors result in larger differences. However, imperfections in virtual brick modeling and lighting estimation can produce significant discrepancies even when the construction is correct. Conversely, small pieces may yield negligible differences even when the user fails to follow the instructions.

To address these limitations, we leverage the IoU for the binary masks of a virtual object's representation and its physical counterpart. As the IoU metric focuses exclusively on location and shape information, our approach redirects challenges such as color inconsistencies, object size variations, and lighting discrepancies into object detection and segmentation, which are well-studied by state-of-the-art ML models. With reasonably accurate detection and segmentation results, the IoU is expected to be high when the MR instructions are correctly followed. At the same time, a low IoU indicates deviations from the intended construction. Additionally, \sysname explicitly accounts for the heterogeneous properties of virtual and physical pieces and leverages the unique characteristics of MR-aided operation applications to facilitate the segmentation of the captured reference and target frames. In the following paragraphs, we detail the auto-verification process of \sysname, including the segmentation methodology and the threshold-based policy that leverages IoU metrics.

\subsubsection{Segment Virtual Pieces in Reference Frames}
In MR applications, virtual objects are created and managed by the system, and their high-level properties, such as center coordinates, shapes, and bounding boxes, are often accessible to users. However, this information alone cannot directly provide a segmentation mask for comparison with the target frame. A straightforward approach would be to project all the vertices of the 3D model onto the screen to generate a perfect mask. However, running such a projection in the main thread becomes computationally prohibitive when the model contains a large number of vertices. Instead, we propose an efficient approach that leverages the rendering pipeline to segment virtual objects from reference frames for further processing. We note that virtual objects in MR applications are rendered on distinct layers from the real-world background. As such, we employ a secondary virtual camera and configure its settings to render only the target virtual elements, as illustrated in \Cref{fig:virtual_info}. This camera renders to a designated texture rather than the user's screen, ensuring no impact on user experience. From the rendered texture, we obtain the segmentation mask by applying a binary filter to identify non-zero pixels on the server side with minimal latency, as shown in \Cref{fig:ref_mask}. An alternative implementation could employ an all-white unlit shader rendered directly to the output texture, further enhancing the system efficiency by eliminating the requirement for applying a binary filter.

\subsubsection{Segment Physical Pieces in Target Frames}
Unlike the readily available properties of virtual objects in the reference frame, segmenting physical objects in the target frame requires both object detection and segmentation. Specifically, the system must first identify whether the target object is present, determine its position, and then generate a segmentation mask. General-purpose segmentation models, such as SAM (see \cite{kirillov2023segany}), are not suitable for our scenario as they segment all objects indiscriminately and lack the contextual understanding needed to identify a specific target object. Instead, we prefer models that integrate object detection with segmentation.
Considering our LEGO implementation and the observation that the number of basic unit categories is significantly smaller than the number of construction steps, we collect a custom dataset and fine-tune a personalized segmentation model based on YOLOv8 (see \cite{Jocher_YOLO_by_Ultralytics_2023}). Note that other segmentation models, such as DUCK-Net (see \cite{dumitru2023using}) and SegFormer (see \cite{xie2021segformer}), are also compatible with our framework. As the design of segmentation models lies outside the scope of this paper, we select YOLOv8 for its widespread accessibility and ease of fine-tuning. An example of segmentation results from our personalized model, including both the masks and classification outputs, is shown in \Cref{fig:seg_results:pmodel}. Given that MR-aided operation applications, such as LEGO assembly, rely on predefined construction paths, we search for the class ID corresponding to the current step within the model outputs to determine whether the target piece exists in the image and, if so, extract its segmentation mask. If multiple objects are detected with the target class ID, we generate multiple target masks and handle them in the next step.
An example of segmentation for the target frame is shown in \Cref{fig:seg_masks:pmodel}.

\subsubsection{Threshold-based Policy Leveraging IoU between Segmentation Results} 
Based on the segmented masks of the reference and target frames, the IoU can be calculated and leveraged in our threshold-based policy. Examples of the segmentation results for the reference and target frames are shown in \Cref{fig:ref_mask} and \Cref{fig:seg_masks:pmodel}, respectively. Multiple target masks may be generated due to duplicate pieces, and in such cases, we select the mask with the maximal IoU value. Finally, we set a threshold based on the empirical analysis of the collected dataset and compare the calculated IoU to this threshold. If the IoU exceeds the threshold, the server determines that the user has successfully built the current bricks and proceeds to the next step by sending corresponding control information to update the virtual models. Otherwise, the user is considered to have failed the current step, and the server waits for the next pair of frames. In \Cref{sec::accuracy}, we analyze the impact of different thresholds on the validation set and identify the threshold with the highest accuracy. 

\subsection{Optimizations for Practical Deployment}\label{sec::practical}
As mentioned earlier, this section outlines our design to address the assumptions required for applying our auto-verification approach, enabling \sysname to function effectively in more practical scenarios.

\subsubsection{Tag-based Localization}
Inaccurate localization significantly affects the performance of the auto-verification process. If the virtual LEGO brick is rendered in the wrong position, a qualified reference frame cannot be obtained. To avoid this, we adopt the open-source framework AprilTag,
known for its low computational overhead and high localization accuracy. Given the pre-defined and fixed relative positions between virtual models and the tag, \sysname can easily localize the virtual model based on the tag's location. Similar to \cite{yan2021augmented,yao2022scalable}, we use small black and white LEGO pieces to build the tag and render corresponding virtual bricks based on their relative directions and distances. Since the LEGO models share the same baseplate as the tag, the virtual bricks maintain the correct position even when we translate or rotate the baseplate, which is common during LEGO construction. In contrast, the localization for a printed tag is unstable due to the shift between the tag and the LEGO model. Note that one tag is sufficient for the LEGO prototype system, whereas multiple tags may be needed for applications involving larger objects.

\begin{figure}[ht]
    \centering
    \subfloat[Reference frame.]{\includegraphics[width=0.15\textwidth]{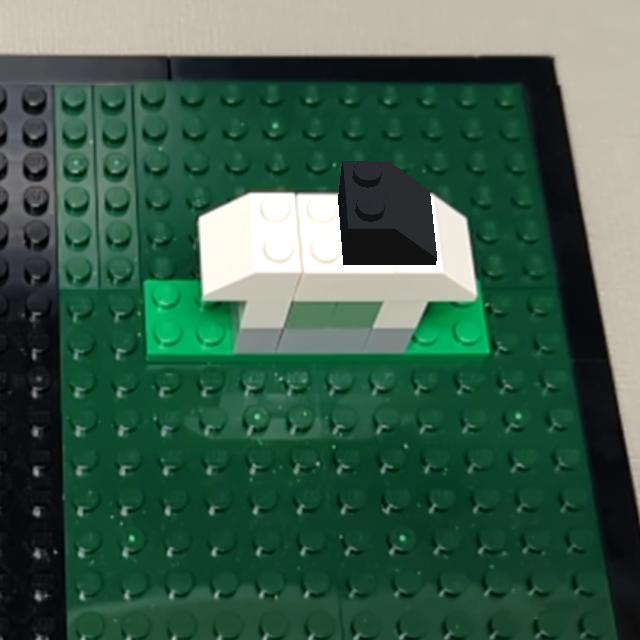}
    \label{fig:ref}}
    \subfloat[Raw target frame.]{\includegraphics[height=0.15\textwidth]{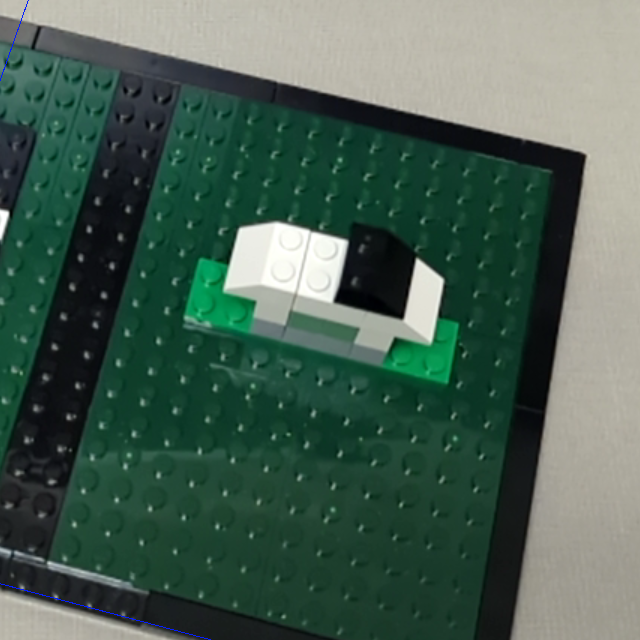}
    \label{fig:tar}}
    \subfloat[Aligned target frame.]{\includegraphics[height=0.15\textwidth]{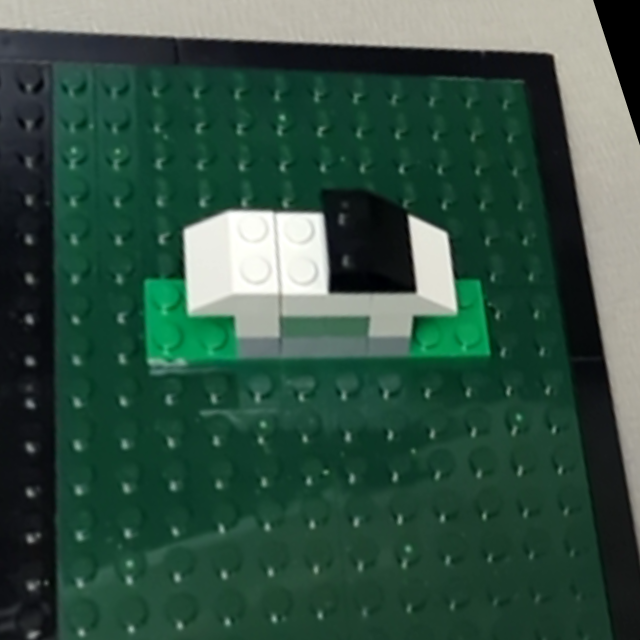}
    \label{fig:tar_align}}
    \caption{Example of frame alignment.
    }
    \label{fig:frame_align}
    \vspace{-0.2in}
\end{figure}

\subsubsection{Frame Alignment} In a real-world scenario, the user is not stationary during interactions with the MR application, leading to potentially low IoU between the reference frame captured before the operation and the target frame captured afterward. Therefore, it is necessary to align the target frame with the reference frame. We propose a method to sample a series of points within the tag coordinate system in both reference and target frames. Then, we leverage their pixel coordinates to calculate the homography matrix and conduct perspective transformation to align the reference and target frame. The aforementioned auto-verification process will be applied to frames after alignment. Note that the sampled points are selected from the corners of the virtual LEGO units to ensure they fall down to the same planar surface.
An example of the frame alignment is shown in \Cref{fig:frame_align}, which demonstrates that our alignment approach is robust to both translational and rotational movements of users.

\subsubsection{Frame Processing} 
Recall that reference and target frames are uploaded to the edge server for further processing. To accelerate the communication between the client and edge server, we crop the frames to remove redundant backgrounds while ensuring the target LEGO bricks remain close to the center. Besides, we downgrade the resolution of frames and apply efficient encoding before uploading to the server.
We define the new resolution
$$[\text{NewWidth},\text{NewHeight}]=\alpha\cdot[\text{CropWidth},\text{CropHeight}],$$
where $\alpha\in(0,1]$ is a scalar parameter. 
We evaluate the impact of $\alpha$ by testing ten different values ranging from $0.1$ to $1$ in \Cref{sec::benchmark::res_impact}. Smaller $\alpha$ values correspond to lower resolution images, which reduce computational latency but decrease verification accuracy.
Without otherwise specified, we set $\alpha=0.5$ in our experiments to balance latency and accuracy. 
Note that the clipping and scaling are performed on the GPU to accelerate the process.

Once the clipping and scaling are finished, the processed image is moved to the memory and compressed by the hardware-accelerated H264 video encoder. Utilizing the hardware-accelerated encoder not only accelerates the encoding process but also significantly reduces energy consumption. The encoded video stream is then uploaded to the edge server for further processing. After receiving the video bytes from the client, the server employs a hardware-accelerated decoder to further reduce the latency. We compare the system performance with PNG, JPG, and H264 video streams and showcase the detailed results in \Cref{sec::benchmark::codec}.
We observe that while H264 and JPG offer lower latency and higher compression compared to PNG, H264 maintains higher accuracy with a slight increment in latency compared with JPG. However, the overall latency is negligible compared to human reaction time, making H264 preferable for MR-aided operations where verification accuracy is more crucial.

\section{System Implementation} \label{sec::imple}
In this section, we illustrate the software and hardware implementation details of \sysname on the client side and the edge server as well as 3D virtual model development. 

\textbf{Client.} The client application of \sysname is developed in Unity by C\# language, which can be exported to commercial Android devices. The localization module adopts the open-source project\footnote{https://github.com/keijiro/jp.keijiro.apriltag} - AprilTag package for Unity. Since copying texture to the CPU is a very expensive operation, we use Unity's built-in \textit{Graphics} API to crop the area and downgrade the resolution directly on the GPU. Particularly, we use \textit{Graphics.CopyTexture()} and \textit{Graphics.Blit()} for clipping and scaling, respectively. Note that \textit{Graphics.Blit()} internally flips the image when processing textures other than the main texture (see \cite{unityshader}). Therefore, we write a shader script to flip the image back to avoid upside-down results. After that, we use \textit{Texture2D.ReadPixels()} to move the data from GPU to CPU. 
We utilize Android MediaCodec (see \cite{mediacodec}) to encode frames into raw H264 video streams and upload them to the server. Note that the main application is developed in Unity, where we cannot directly deploy the MediaCodec. As such, we develop a wrapper for the MeidaCodec as an Android library and export it to an AAR file, which can be imported as a plugin in Unity.

\textbf{Edge Server.} The server of \sysname is written in Python, which works in either Windows or Linux systems. Most of the computer vision technologies use existing APIs from \emph{NumPy} and \emph{scikit-image}.
To accelerate the decoding of the video stream, we develop a dynamic link library (DLL) to utilize NVDEC in C++ based on NVIDIA Video Codec SDK (see \cite{videocodec}). We then write a wrapper in Python to call this library.
The server and the client communicate over Transmission Control Protocol (TCP).
To facilitate the full-duplex data transmission between the server and the client, we write an application layer communication protocol
to handle uploading the frames and downloading the detection results. 

\begin{figure}[ht]
    \vspace{-0.1in}
    \centering
    \subfloat[Physical LEGO models.]{\includegraphics[height=0.18\textwidth]{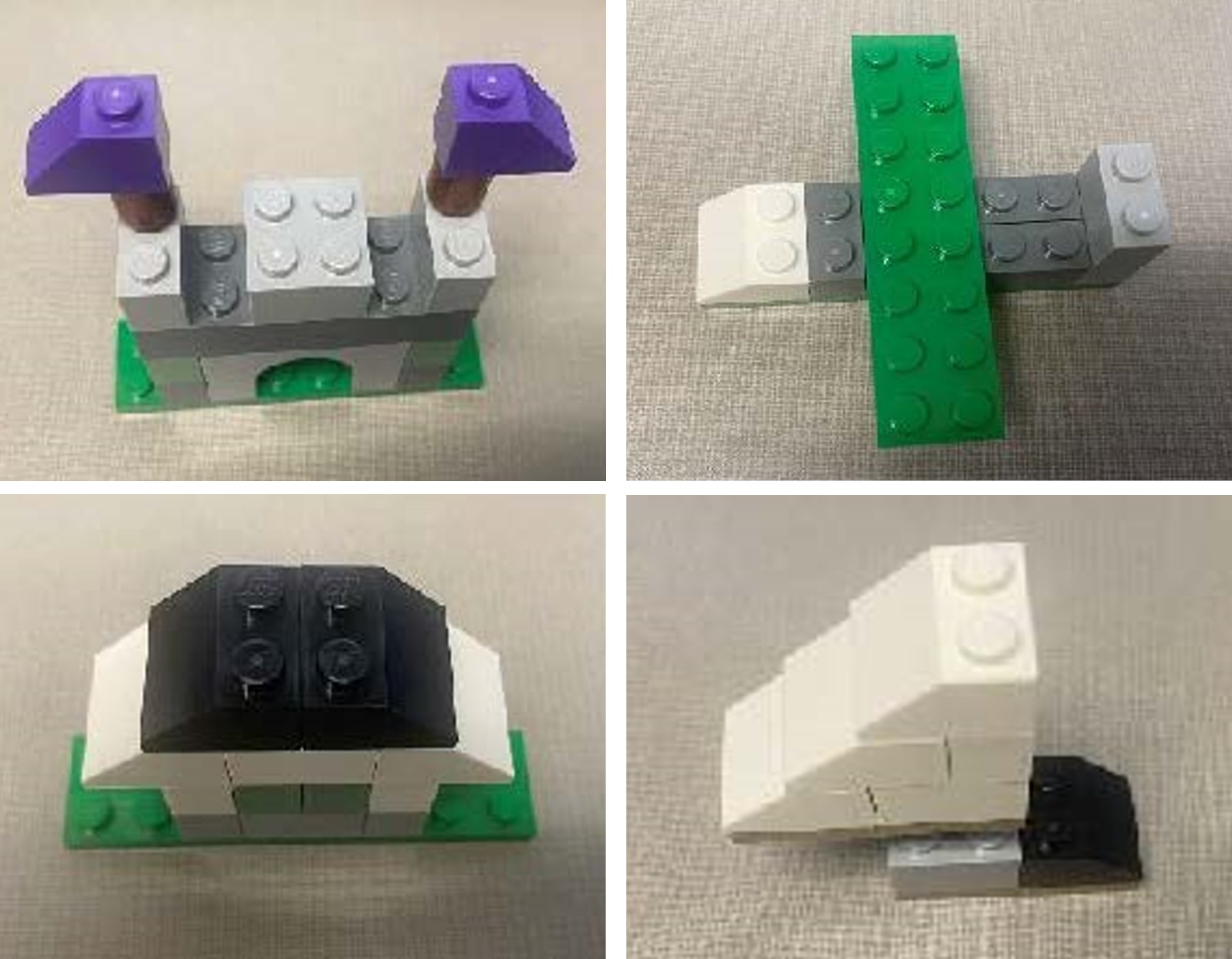}}
    \hfill
    \subfloat[Virtual LEGO models.]{\includegraphics[height=0.18\textwidth]{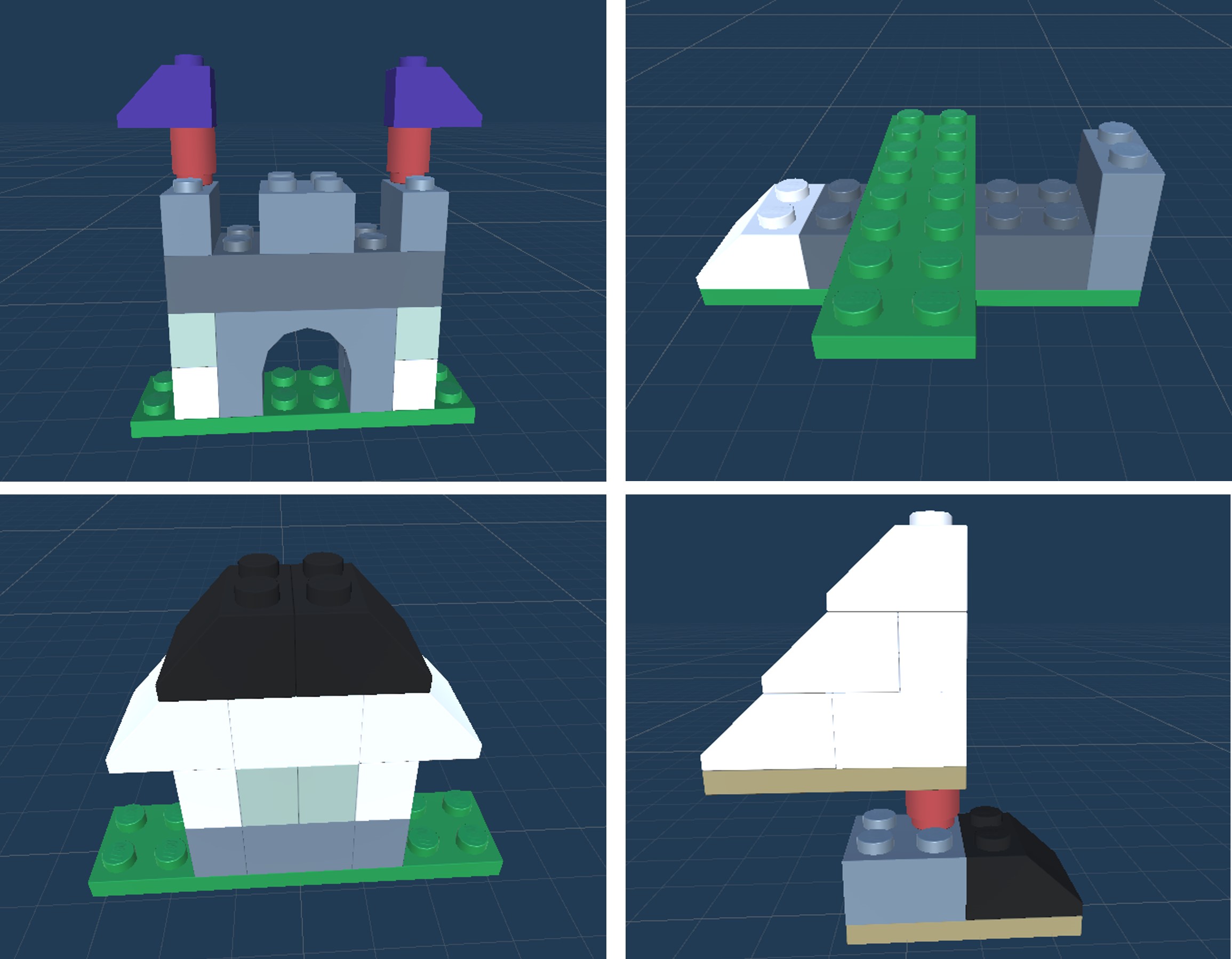}}
    \caption{LEGO models used in experiments.}
    \label{fig:lego-model}
    \vspace{-0.1in}
\end{figure}

\textbf{LEGO Model.} 
We use the LEGO Classic Creative Suitcase 10713\footnote{https://www.lego.com/en-us/product/creative-suitcase-10713} 
and select four different LEGO models that can be built from shared LEGO pieces, as shown in \Cref{fig:lego-model}.
\sysname can be easily applied to other operation applications as long as the 3D models are available. To acquire a virtual model that can be rendered as MR objects, we use BrickLink Studio to build our target LEGO model using official LEGO bricks. Then, the model is exported to the Unity project via LEGO Model Importer (see \cite{legoimporter}).

\begin{figure}[ht]
    \centering
    \includegraphics[width=0.48\textwidth]{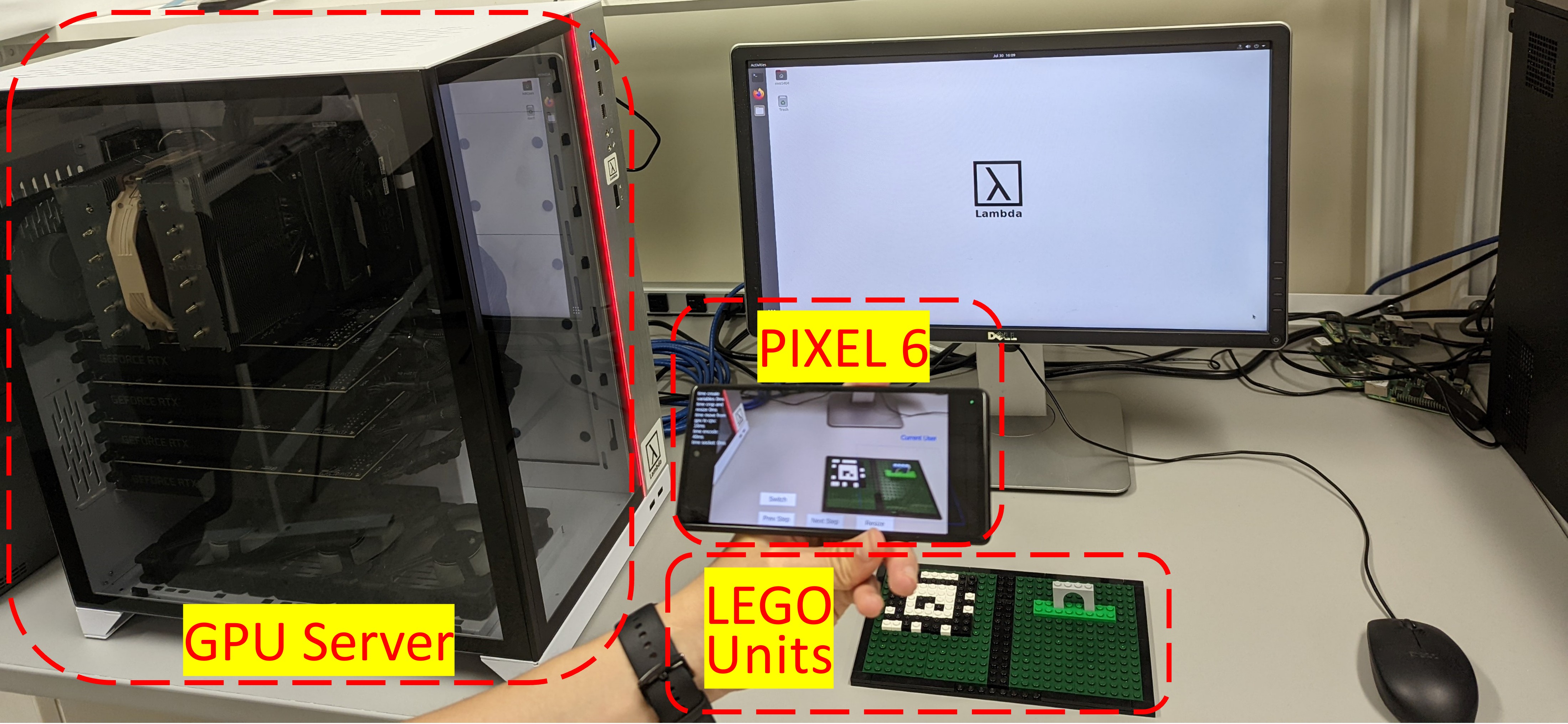}
    \caption{Prototype of our system.
    }
    \label{fig:prototype}
    \vspace{-0.2in}
\end{figure}

We deploy the client application of \sysname on a Google Pixel 6 with Android 13. 
The server is deployed on a Lambda workstation with Intel Core i9-10980XE CPU @3.00GHz × 36, NVIDIA GeForce RTX 3070 Graphics Card × 4, 128 GB memory, 4 TB disk, and Ubuntu 20.04 LTS. The communication between the edge server and the client is handled by a TP-Link Archer AX50 router, with the server connected via a wired cable and the client connected wirelessly. The experiments are conducted in a typical office room, where the mobile device, server, and router are located. The smartphone is held by the user to emulate the movement of MR headsets.
The prototype is shown in \Cref{fig:prototype}.

\section{Evaluations} \label{sec::eva}
In this section, we first generate large-scale synthetic datasets based on public datasets and collect our dataset based on the LEGO prototype.
Next, we evaluate the classification accuracy and end-to-end latency of our system compared to several baseline approaches. Besides, we measure the computational overhead and energy consumption of \sysname compared to the MR-aided operation system without auto-verification.

\subsection{Dataset Preparation and Analysis}
To comprehensively evaluate the proposed auto-verification policy, we use datasets from various sources. Given the absence of public datasets directly related to MR-aided operation contexts, we select two public segmentation datasets associated with PCB boards (see \cite{makwana2023pcbsegclassnet}, also known as FICS PCB Image Collection (FPIC) Component Dataset) and breadboards\footnote{ https://universe.roboflow.com/cv2023/simple-circuit}
to simulate laboratory operations environments in MR contexts. The PCB board dataset includes 5008 training and 1252 validation samples across 25 classes, while the breadboard dataset contains 261 training and 21 validation samples within 3 classes.
It's worth noting that we fine-tune two YOLO models using the training samples from those two datasets for accurate segmentation and classification of target frames. 

Besides fine-tuning the YOLO model, we process the dataset to fit the auto-verification tasks in MR-aided operation contexts. For each dataset image, we extract ground truth labels, including segmentation masks and corresponding classification outcomes, to create reference and target frames. 
We designate original images as target frames and create corresponding reference frames by simulating a 3D virtual model effect, i.e., applying multiple filters to areas defined by each segmentation mask in the labels. To generate samples representing correct operations, we overlay processed areas directly onto the specified region. For incorrect operation samples, we apply a random shift to each region using a uniform distribution ranging from 0.5 to 1 times the bounding box's side length on each axis. 
After the conversion process, the validation set of the PCB board dataset comprises 9,752 samples, while the test set contains 9,796 samples. The breadboard dataset includes 248 validation samples and 204 test samples. In each set, half of the samples represent correct operations.

\begin{figure}[ht]
  \vspace{-0.2in}
  \centering
  \subfloat[Dataset collection.]{
    \includegraphics[height=0.12\textwidth]{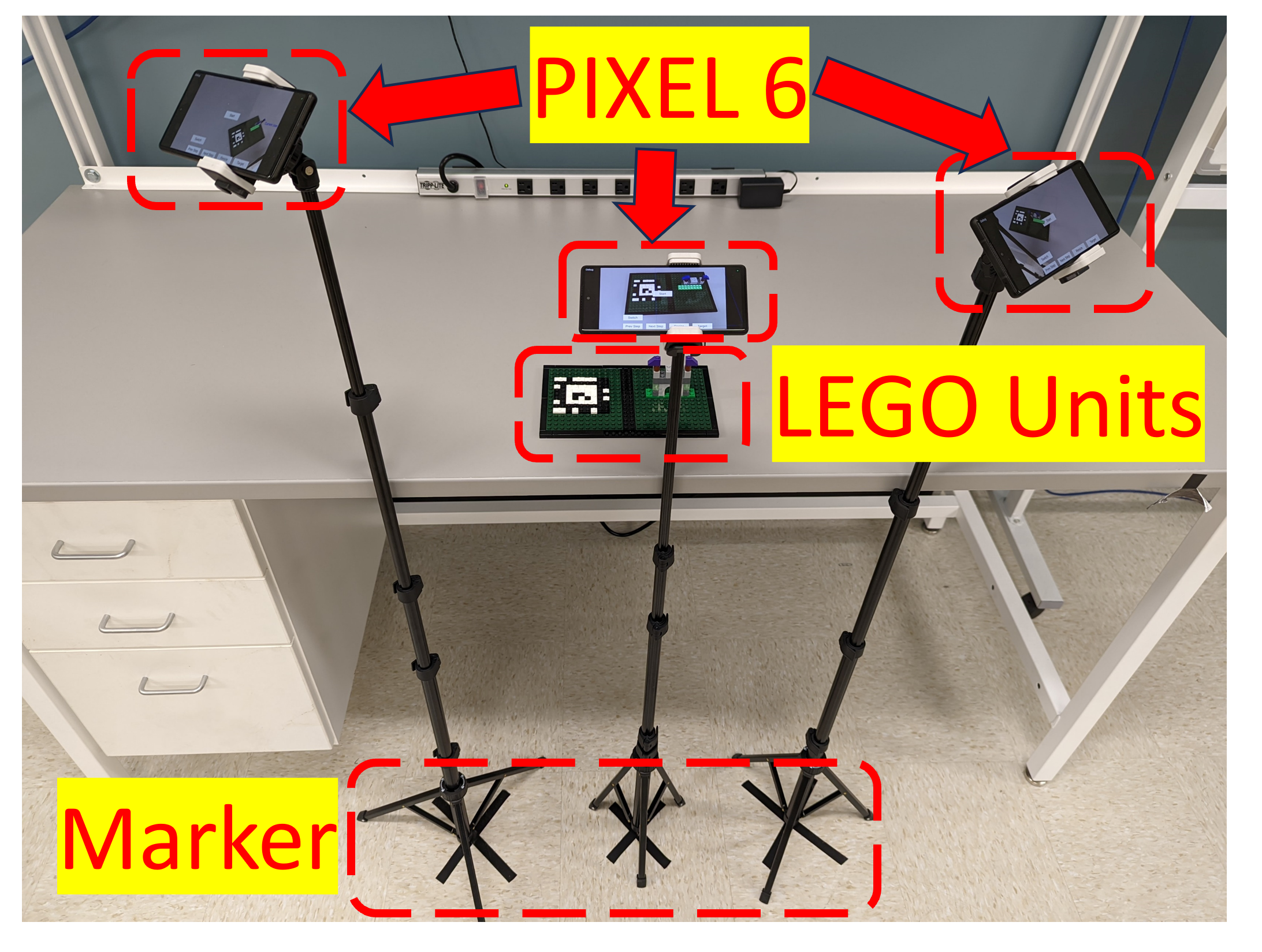}
    \label{fig:dataset_collection}
    }
  \subfloat[Annotated frame.]{
    \includegraphics[height=0.12\textwidth]{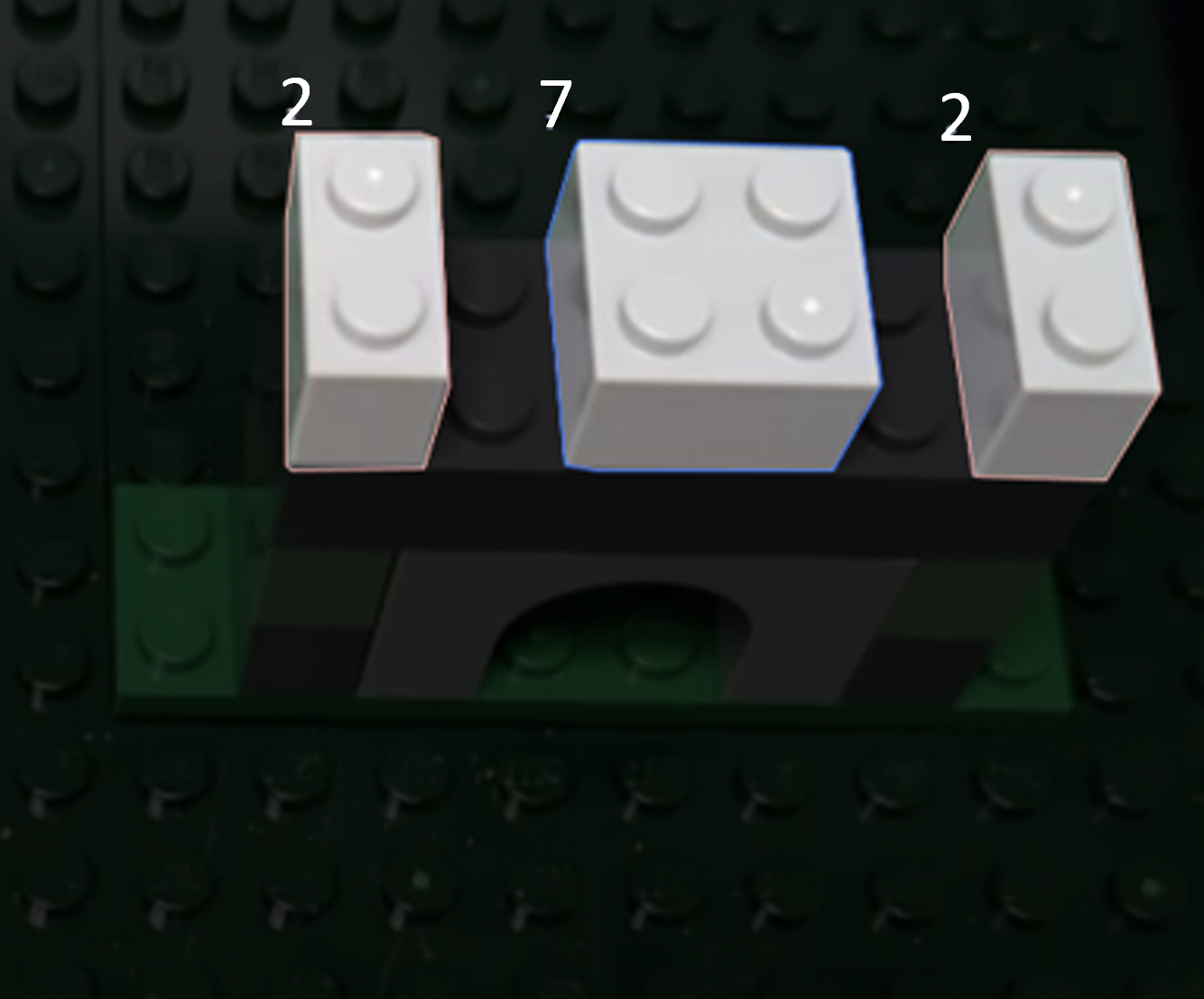}
    \label{fig:ex-targetframe-anno}
  }
  \subfloat[Learning curve.]{
    \includegraphics[height=0.12\textwidth]{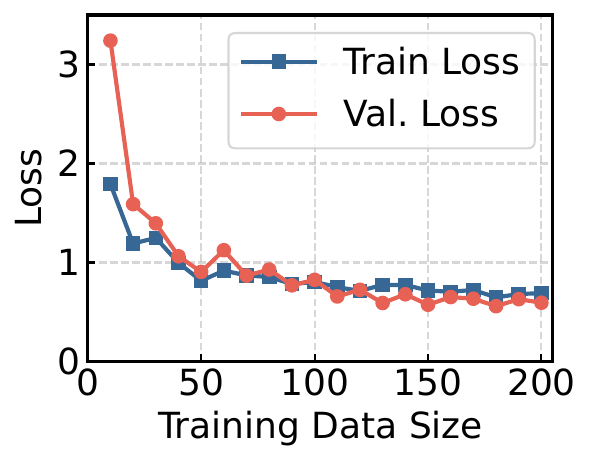}
    \label{fig:learncurve}
  }
  \caption{Custom dataset preparation and training.
  }
\end{figure}

In addition to the synthetic datasets, we develop our own dataset based on our system implementation. As shown in \Cref{fig:dataset_collection}, we use a Google Pixel 6 to capture data, marking three distinct locations on the smartphone holders with a black marker. These holders are set at three different heights: 115 cm, 95 cm, and 105 cm, from left to right. Furthermore, we utilize LEGO baseplates of various colors to enhance dataset diversity. This systematic dataset collection process ensures the robustness of our fine-tuned model. All frames were saved as PNG files to prevent data loss from compression, with an image resolution set to 640x640 pixels. For efficient segmentation and labeling, we employ Roboflow (see \cite{dwyer2022roboflow}), a tool that complements our workflow by allowing us to annotate each LEGO piece in the target frames with a specific class ID. An example of the annotation is shown in \Cref{fig:ex-targetframe-anno}. 

Our custom LEGO dataset contains 200 samples for fine-tuning the YOLO model, 74 samples for validating the tuning results, and 500 samples each for our validation and test sets. It covers all four models with 46 steps and includes 15 unique piece classes.
Each sample in the validation and test sets includes a pair of raw RGB images for reference and target frames, the model ID, and the step index. 
Half of the samples depict the user correctly following the virtual instructions and accurately positioning the LEGO bricks. The other half simulates incorrect actions, with pieces randomly placed at different stages of the assembly process.

To verify the sufficiency of the number of samples in our dataset, we draw the learning curve, i.e., the training loss and validation loss against the number of samples used for training, as shown in \Cref{fig:learncurve}, where the average train/validation loss refers to the average of bounding box loss, segmentation loss, classification loss and distribution focal loss on the train/validation set. 
The training loss reflects how effectively the model learns from varying amounts of data, while the validation loss indicates how generalization improves as the training set grows.
From the learning curve, we can see both training and validation losses decreasing to a stable point with minimal disparity between their final values, implying that our dataset is sufficiently large to support generalizable learning.
In the next section, we will compare the classification accuracy of \sysname with several baseline approaches to demonstrate its efficacy, which also reveals that we can achieve favorable results with such a small number of samples used for fine-tuning the model.

\subsection{Auto-Verification Accuracy Comparison}
\label{sec::accuracy}
In this section, we compare three different automatic verification methods to demonstrate the efficiency of \sysname: (1) \sysname, (2) Sim-Ver, and (3) ML-Ver.
Sim-Ver (see \cite{yao2023design}) directly calculates the traditional similarity metrics based on the reference and target frames and utilizes those metrics to do auto-verification. Here, we use the PSNR because it outperforms other metrics like SSIM, normalized root mean square error (NRMSE), and normalized cross-correlation (NCC). The detailed comparison results are presented in \Cref{sec::benchmark::sim}.
As for ML-Ver (revised from \cite{alshowaish2022trademark} and \cite{simvgg16}), we deploy a pre-trained YOLOv8 classification model to process the reference and target frames. The cosine similarity between the outputs, i.e., embeddings that represent the possibility for each class, is used to classify whether the user correctly follows the instructions.
Note that the main difference between ML-Ver and \sysname lies in the output of the ML models and the metrics used for verification.

\begin{figure*}[t]
    \centering
    \begin{minipage}[t]{0.58\textwidth}
        \centering
        \subfloat[ROC - PCB board.]{
            \includegraphics[width=0.32\linewidth]{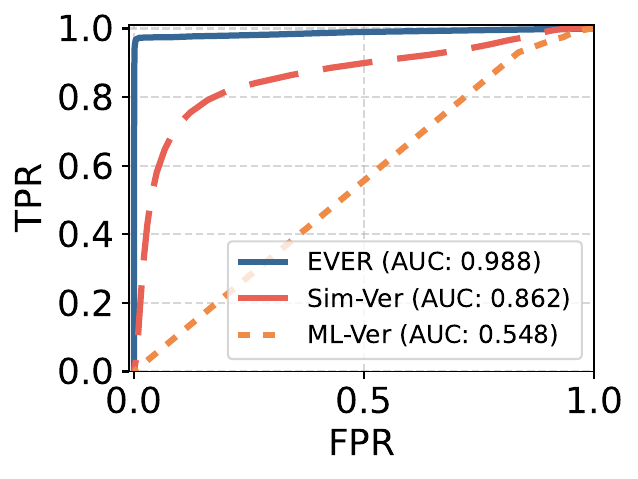}
            \label{fig:ROC_fpic}
        }
        \subfloat[ROC - breadboard.]{
            \includegraphics[width=0.32\linewidth]{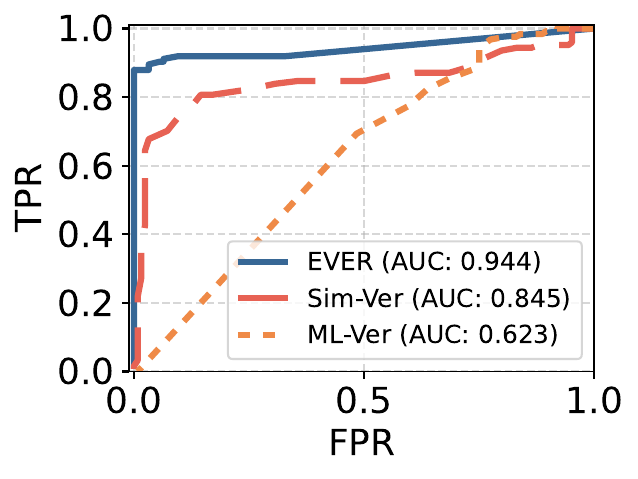}
            \label{fig:ROC_bread}
        }
        \subfloat[ROC - LEGO.]{
            \includegraphics[width=0.32\linewidth]{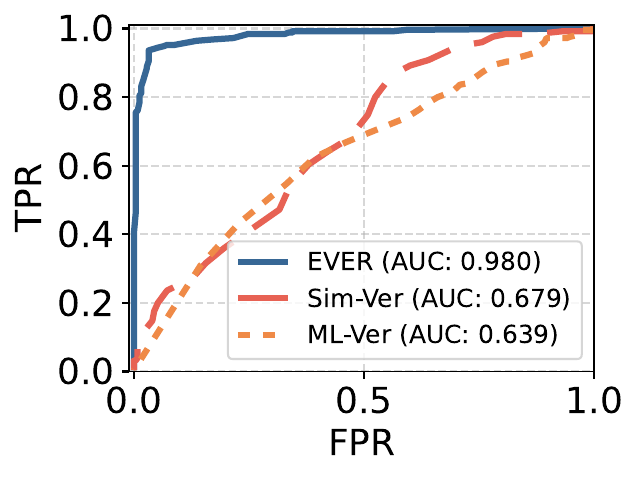}
            \label{fig:ROC_lego}
        }
        
        \subfloat[ACC - PCB board.]{
            \includegraphics[width=0.32\linewidth]{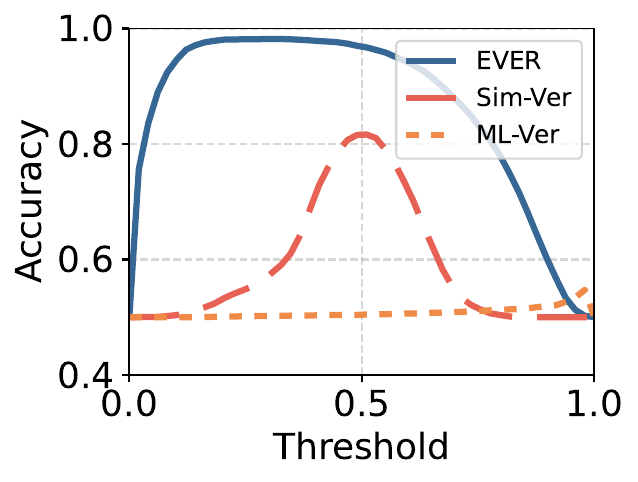}
            \label{fig:aucvalid_fpic}
        }
        \subfloat[ACC - breadboard.]{
            \includegraphics[width=0.32\linewidth]{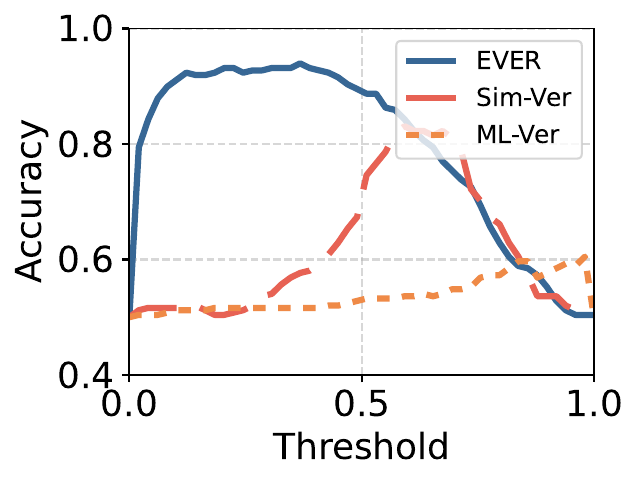}
            \label{fig:aucvalid_bread}
        }
        \subfloat[ACC - LEGO.]{
            \includegraphics[width=0.32\linewidth]{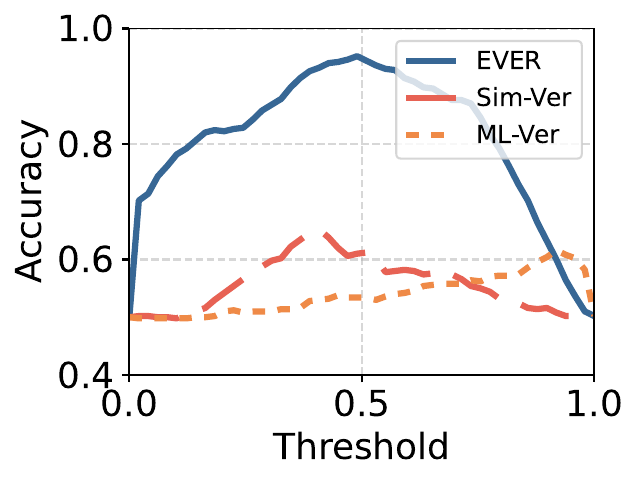}
            \label{fig:aucvalid_lego}
        }
    
        \caption{Comparison of ROC and ACC results on the validation set.}
        \label{fig:auc_acc_all}
    \end{minipage}
    \hspace{0.02\textwidth}
    \begin{minipage}[t]{0.38\textwidth} 
        \centering
        \subfloat[End-to-end latency.]{%
            \includegraphics[width=0.48\linewidth]{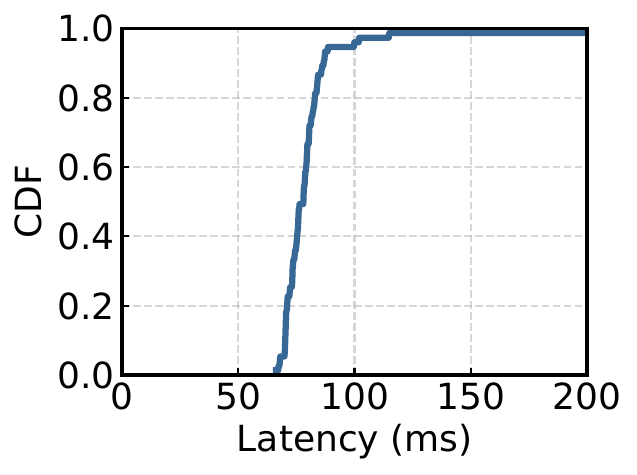}%
            \label{fig:lat_valid}%
        }
        \hfill
        \subfloat[Inference latency.]{%
            \includegraphics[width=0.48\linewidth]{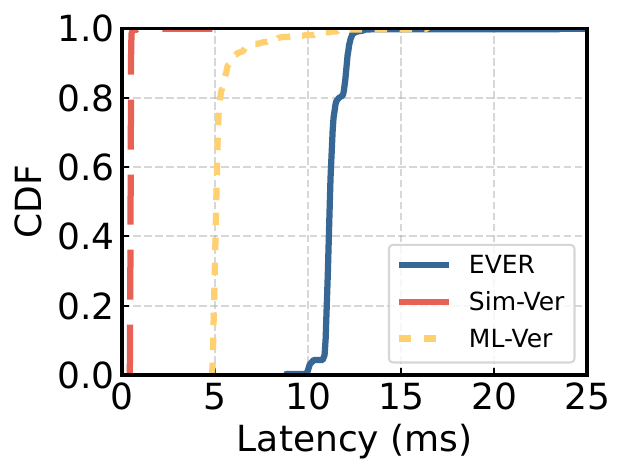}%
            \label{fig:lat_comp}%
        }
        
        \subfloat[Size of frames.]{%
            \includegraphics[width=0.48\linewidth]{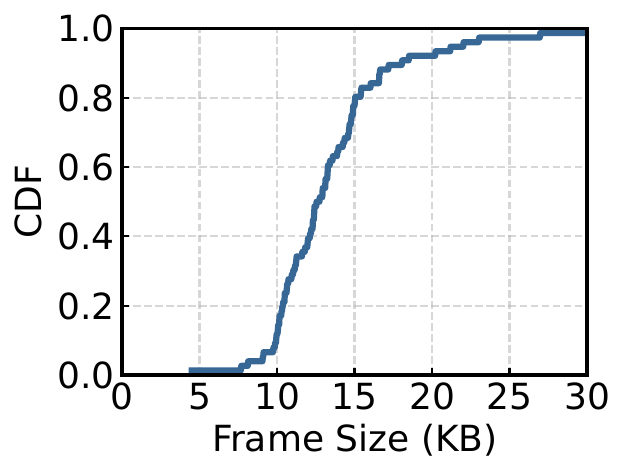}%
            \label{fig:size_frame}%
        }
        \hfill
        \subfloat[CPU usage (max:800\%).]{%
            \includegraphics[width=0.48\linewidth]{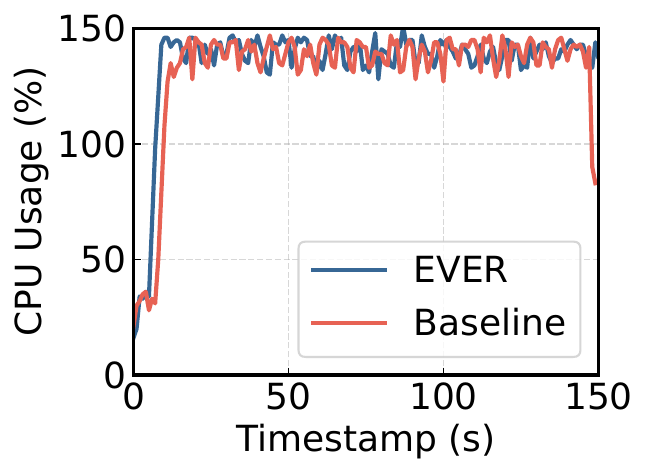}%
            \label{fig:cpu_usage}%
        }
        
        \caption{Real-time system evaluation.}
        \label{fig:real-sys}
    \end{minipage}
    \vspace{-0.2in}
\end{figure*}

We define a \emph{positive event} and a \emph{negative event} as the pair of frames when the user correctly and erroneously follows the instruction, respectively. We have a \emph{positive detection} if the metric is higher than the threshold. In contrast, a metric lower than the threshold implies a \emph{negative detection}. Note that for any verification task, there are four possible results: True Positive (TP) and False Negative (FN) stand for a positive event with positive detection and negative detection, respectively; False Positive (FP) and True Negative (TN) represent a negative event with positive detection and negative detection, respectively. 
To assess the effectiveness of our auto-verification policy, we focus on the following metrics: Positive Predicted Value (PPV, or precision), which is the proportion of the accurate verification within the positive detections; True Positive Rate (TPR, or recall/sensitivity), representing the fraction of accurately verified instances out of all positive events; False Positive Rate (FPR, or fall-out), indicating the proportion of erroneous verifications among all negative events; and Accuracy (ACC), which measures the number of accurate verification results in the entire dataset. Our goal is to maximize PPV, TPR, and ACC while minimizing the FPR.

We first draw the Receiver Operator Characteristic (ROC) curve of those three policies on three datasets in \Cref{fig:auc_acc_all} (a)-(c), illustrating the trade-off between TPR and FPR with different threshold setups. Specifically, we apply various thresholds to the validation set to get the corresponding FPR, TPR, and ACC results. 
From the ROC figure, we can calculate the Area Under the ROC Curve (AUC) and observe that our approach achieves the best AUC among all three datasets.
We also plot the verification accuracy against the thresholds to assess model robustness, as shown in \Cref{fig:auc_acc_all} (d)-(f). The values of IoU, PSNR, and cosine similarity used in those three approaches are normalized to the range $(0,1)$ for better visualization. The results demonstrate that our approach achieves much better ACC compared to the two baseline methods.
Besides, the performance drop of Sim-Ver in our custom dataset indicates its sensitivity to camera vibrations during data collection. 

\begin{table}[hbtp]
\centering
\begin{tabular}{ccccc}
\toprule
\textbf{Metric} & \textbf{EVER} & \textbf{SIM-Ver} & \textbf{ML-Ver} & \textbf{Dataset} \\
\midrule
\textbf{PPV} & \textbf{0.9920} & 0.8751 & 0.5286 & \multirow{4}{*}{PCB} \\
\textbf{TPR} & 0.9149 & 0.7438 & \textbf{0.9269} &  \\
\textbf{FPR} & \textbf{0.0073} & 0.1062 & 0.8265 &  \\
\textbf{ACC} & \textbf{0.9538} & 0.8188 & 0.5502 &  \\
\midrule
\textbf{PPV} & \textbf{0.9773} & 0.7949 & 0.5398 & \multirow{4}{*}{Breadboard} \\
\textbf{TPR} & \textbf{0.8431} & 0.6078 & 0.5980 &  \\
\textbf{FPR} & \textbf{0.0196} & 0.1569 & 0.5098 &  \\
\textbf{ACC} & \textbf{0.9118} & 0.7255 & 0.5441 &  \\
\midrule
\textbf{PPV} & \textbf{0.9615} & 0.5915 & 0.5983 & \multirow{4}{*}{LEGO} \\
\textbf{TPR} & \textbf{0.9000} & 0.8920 & 0.5480 &  \\
\textbf{FPR} & \textbf{0.0360} & 0.6160 & 0.3680 &  \\
\textbf{ACC} & \textbf{0.9320} & 0.6380 & 0.5900 &  \\
\bottomrule
\end{tabular}
\caption{{Evaluation metrics on the test set.}}
\label{fig:testConfusionresult}
\vspace{-0.1in}
\end{table}

Given the optimal threshold obtained via ROC and ACC curves, we present the evaluation metrics for each approach on the test set in \Cref{fig:testConfusionresult}.
\sysname achieves the highest PPV, TPR, and ACC (over 90\% among all datasets) while maintaining the lowest FPR in most cases. 
Moreover, the poor results of Sim-Ver and ML-Ver indicate that traditional similarity-based approaches, whether reliant on image quality metrics or embeddings from machine learning models, lack the necessary sensitivity for auto-verification tasks within MR-aided operation contexts. This deficiency primarily stems from these methods' failure to account for the discrepancies between virtual modeling and physical objects.

\subsection{End-to-End Latency}
In this section, we evaluate the real-time performance of the proposed approach in our edge-assisted system. The three approaches mentioned in \Cref{sec::accuracy} share the same framework, differing only in the server-side inference process. Therefore, we focus on assessing the end-to-end latency of our approach and compare the inference latency with baselines.

In our system, end-to-end latency refers to the time between the user's motion detected until the mobile device receives the verification results from the server and displays them on the screen,
which is measured for each operation. The lower the latency, the more responsive the system is. We expect that the latency is lower than the human reaction time, i.e., $273$ ms \cite{humanreaction}. We run \sysname for five minutes and emulate the construction of LEGO models to measure the latency. The CDF of end-to-end latency is shown in \Cref{fig:lat_valid}. Specifically, the mean and median are $80.13$ ms and $78.03$ ms, respectively, which are far lower than the human reaction time. We also compare the inference latency of three approaches and draw the CDF of the latency in \Cref{fig:lat_comp}. We run those three methods on all three datasets, including validation and test sets, and record the inference latency for each auto-verification step.
The mean inference latencies for the three approaches are $16.17$ ms, $1.37$ ms, and $5.61$ ms, respectively. Although \sysname has a higher inference latency than the baselines, the latency remains significantly low considering the human reaction time.

\begin{figure*}[t] 
    \centering
    \begin{minipage}[hbtp]{0.25\textwidth}
        \centering
        \includegraphics[width=\linewidth]{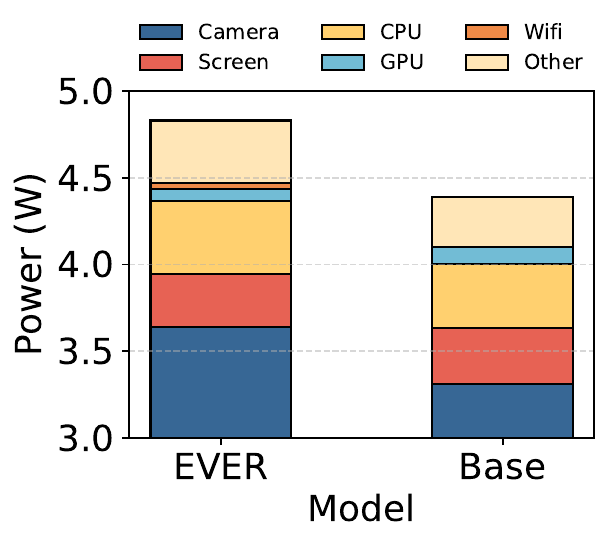}
        \caption{Power consumption (W).}
        \label{fig:power-component}
    \end{minipage}
    \hfill
    \begin{minipage}[hbtp]{0.7\textwidth} 
        \centering
        \subfloat[Latency.]{%
            \includegraphics[width=0.32\linewidth]{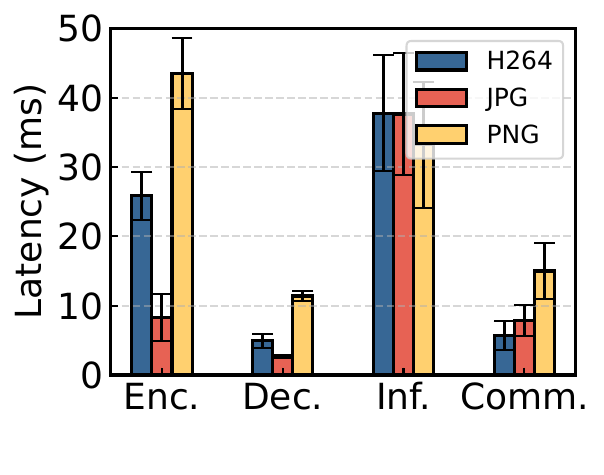}%
            \label{fig:lat_metric}%
        }
        \hfill
        \subfloat[Frame size.]{%
            \includegraphics[width=0.32\linewidth]{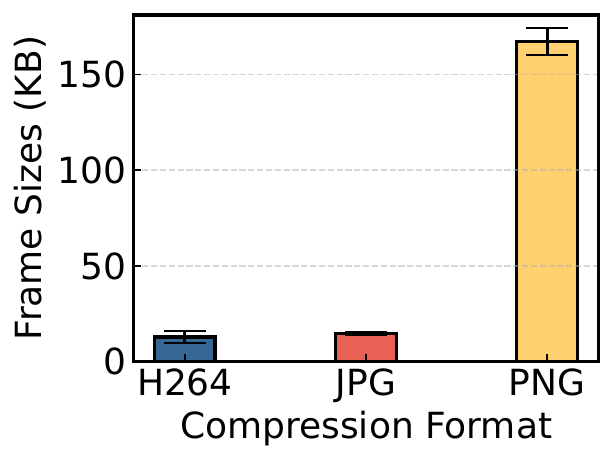}%
            \label{fig:frame_metrics}%
        }
        \hfill
        \subfloat[Accuracy.]{%
            \includegraphics[width=0.32\linewidth]{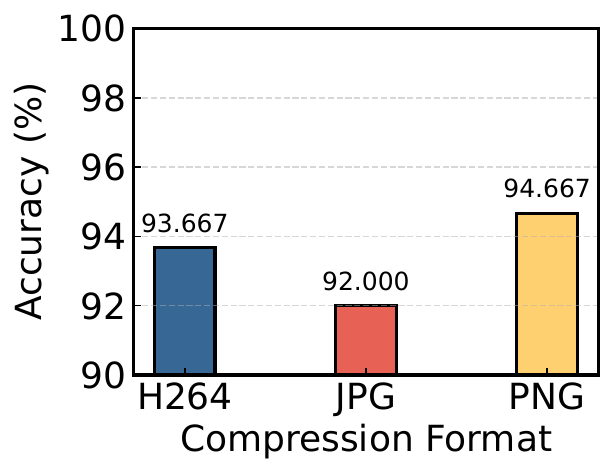}%
            \label{fig:acc_metrics}%
        }
        \caption{Impact of Hardware-Accelerated Video Codec.}
        \label{fig:codec_metrics}
    \end{minipage}
    \vspace{-0.2in}
\end{figure*}

On the other hand, we measure the frame rate of our system in a 1-second sliding window. The target frame rate is set to 60 frames per second (FPS) to ensure a smooth display of virtual content and real-world environment. Note that the rendering pipeline runs in parallel with the auto-verification process. Thus, a larger end-to-end latency doesn't necessarily mean a lower frame rate.
The evaluation results show that the mean and median frame rate are $60.79$ FPS and $61$ FPS, respectively. Furthermore, we collect the size of reference and target frames and draw the CDF in \Cref{fig:size_frame}. With the help of the proposed hardware-accelerated encoding, the average frame size is only 13.42 KB, reducing the frame size by $99.2\%$ compared with the raw RGB texture of $640\times640$ resolution. 

To further understand the latency components of our system, we also collect the detailed latency for each part of the auto-verification process. Notably, it includes frame pre-processing, frame encoding on the client, frame decoding on the server, frame post-processing, and wireless communication. Recall that pre-processing contains image cropping, scaling on the GPU, and moving the image data from the GPU to the CPU. Post-processing refers to the segmentation of the target frames and the calculation of IoU, together with some IO operations to fit YOLO model inputs.
Note that the segmentation for the reference frames is not counted as it can be completed during the user's action, which typically takes several seconds.
As shown in \Cref{fig:lat_detail}, the post-processing of the frames is the main component of the latency, as it requires real-time inference using the machine learning model. The encoding, decoding, and communication latency could be further improved by reducing the resolution of the uploaded image, with a sacrifice of lower accuracy.

\begin{table}[h!]
\centering
\begin{tabular}{@{\hskip 2pt}c@{\hskip 2pt}c@{\hskip 2pt}c@{\hskip 2pt}c@{\hskip 2pt}c@{\hskip 2pt}c@{\hskip 2pt}c@{\hskip 2pt}}
\toprule
\textbf{Stage} & Pre-proc. & Encoding & Decoding & Post-proc. & Comm. \\
\midrule
\textbf{Latency (ms)} & 9.42  & 17.04  & 5.81  & 40.96  & 6.91  \\
\bottomrule
\end{tabular}
\caption{Latency decomposition.}
\label{fig:lat_detail}
\vspace{-0.2in}
\end{table}

\subsection{Resource and Energy Usage} 

To demonstrate the lightweight property of our edge-assisted system, we measure the CPU usage and energy consumption during runtime using the ADB tools and Batterystats. 
The profiling of our application shows an average CPU usage of $135.938$\% and memory consumption of $510.724$ MB. We also profile a baseline scenario where auto-verification is disabled, which records a CPU usage of $131.833$\% and a memory usage of $485.267$ MB. This comparison, as shown in \Cref{fig:cpu_usage}, reveals that integrating auto-verification into our system results in a marginal increment in resource demands – approximately a $4$\% rise in CPU usage and an additional $30$ MB of memory.
Note that the reported CPU usage values greater than $100$\% stem from multi-core utilization. The Google Pixel 6 device used in our experiments has 8 CPU cores, meaning that the maximum theoretical CPU usage can reach $800$\%. Thus, our system utilizes less than $20$\% of total CPU capacity, indicating efficient resource usage.
For energy profiling, we calculate the total power consumption over the duration of a representative MR-guided task, which lasted approximately 30 minutes, and then normalize the power usage over the actual runtime. This normalization enables fair comparison across tasks with different durations. Our system's average power stands at $4.831$ watts.
The detailed power statistics for each component of \sysname and the baseline are shown in \Cref{fig:power-component}. The most energy-intensive component is the camera, which consumes 3.64 watts and is necessary for all MR applications. We observe that the statistics of \sysname are quite close to the baseline without auto-verification. This implies that our verification process is quite lightweight and practical to run on mobile devices. 

\subsection{Benchmarking}
In this section, we present several benchmarks to demonstrate the impact of our key design decisions. 

\subsubsection{Impact of Hardware-Accelerated Video Codec}\label{sec::benchmark::codec}
To study the impact of hardware-accelerated video codec, we test the performance with PNG, JPG, and H264 video streams using our custom dataset to guarantee a fair comparison between those strategies. Specifically, we store the dataset in the hard disk of the mobile device and utilize those frames as \sysname's inputs.
Meanwhile, the encoder and the decoder are updated to be consistent with those codecs.
The average latencies and other related metrics are shown in \Cref{fig:codec_metrics}. Recall that our dataset is collected and stored as PNG files due to the lossless compression. Compared with PNG, JPG and H264 provide less encoding/decoding latency and a much higher compression ratio with a sacrifice of image quality. Specifically, the average frame sizes of H264 and JPG are only $7.65\%$ and $8.77\%$ of the PNG files, respectively. While both H264 and JPG formats lead to accuracy decrements, H264 yields a higher accuracy than the JPG format, suggesting that H264 offers better image qualities. Although the encoding and decoding latency of JPG is the lowest, we note that MR-aided operation applications are more sensitive to verification accuracy than latency. 

\subsubsection{Impact of Resolution Downscaling}
\label{sec::benchmark::res_impact}
\begin{figure}[htb]
    \centering
    \subfloat[Latency impact.]{%
        \includegraphics[height=0.37\linewidth]{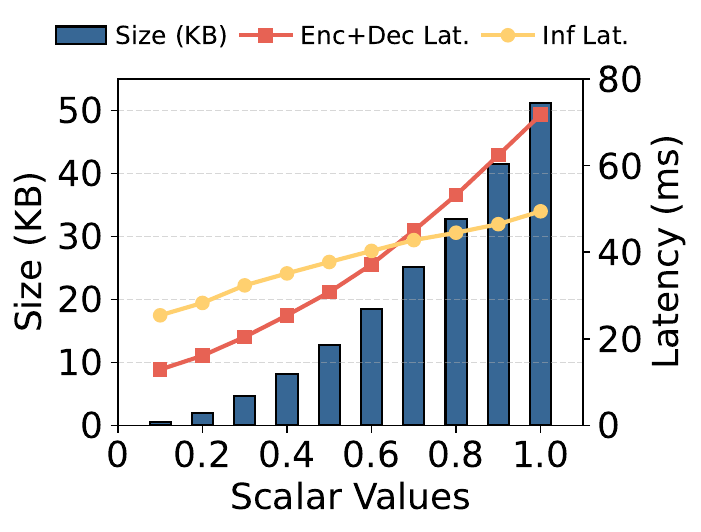}%
        \label{fig:sca_lat}%
    }
    \hfill
    \subfloat[Accuracy impact.]{%
        \includegraphics[height=0.37\linewidth]{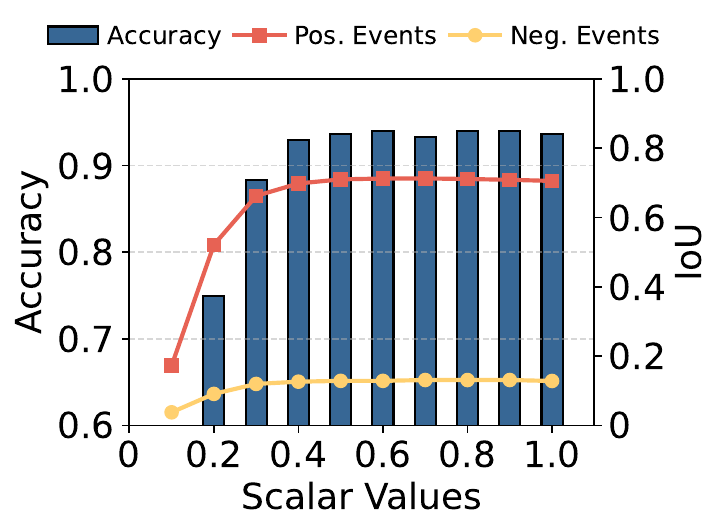}%
        \label{fig:sca_iou}%
    }
    \caption{Impact of scalar values.}
    \label{fig:simtime}
\end{figure}

As mentioned in \Cref{sec::design}, we downgrade the resolution of the cropped image to further reduce the computational overhead. Recall that we define a scalar value $\alpha$ to serve as the scaling factor to adjust the image resolution. The larger the $\alpha$, the higher the image resolution. To study the impact of $\alpha$ on the computation latency and the verification accuracy, we evaluate ten different $\alpha$ values from $0.1$ to $1$. Intuitively, an image with a lower resolution incurs less latency, as the encoding, decoding, communication, and inference time should be smaller. In contrast, lower resolution makes it difficult to differentiate between positive and negative events. As shown in \Cref{fig:sca_lat}, the computation latencies and frame sizes drop significantly with the decrement of the $\alpha$ value. On the other hand, \Cref{fig:sca_iou} shows that when $\alpha$ is too small, we can see that the IoU of the positive and negative events are close to each other, corresponding to a lower verification accuracy. As such, we set $\alpha=0.5$ to significantly reduce the latency without significantly affecting the verification accuracy.

\vspace{-0.2in}

\vspace{-0.1in}
\subsubsection{Impact of Different Similarity Metrics} \label{sec::benchmark::sim}
\begin{figure}[htb]
    \centering
    \subfloat[CDF of PSNR values.]{%
        \includegraphics[width=0.49\linewidth]{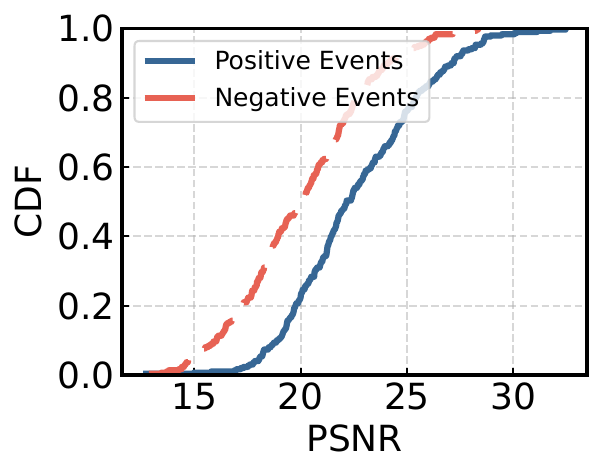}%
        \label{fig:psnr}%
    }
    \hfill
    \subfloat[CDF of NRMSE values.]{%
        \includegraphics[width=0.49\linewidth]{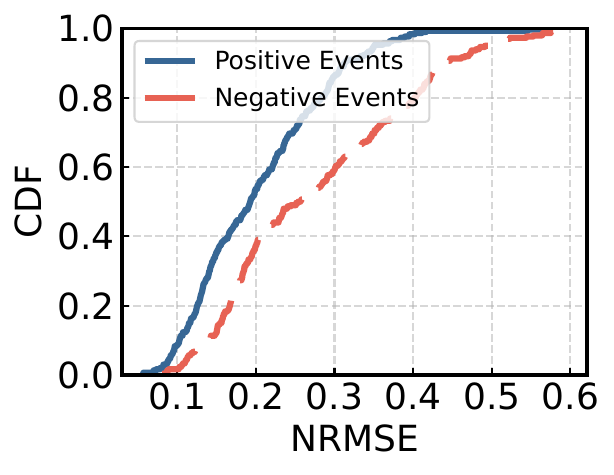}%
        \label{fig:cdfrmse}%
    }
    \vspace{-0.1in}
    \subfloat[CDF of SSIM values.]{%
        \includegraphics[width=0.49\linewidth]{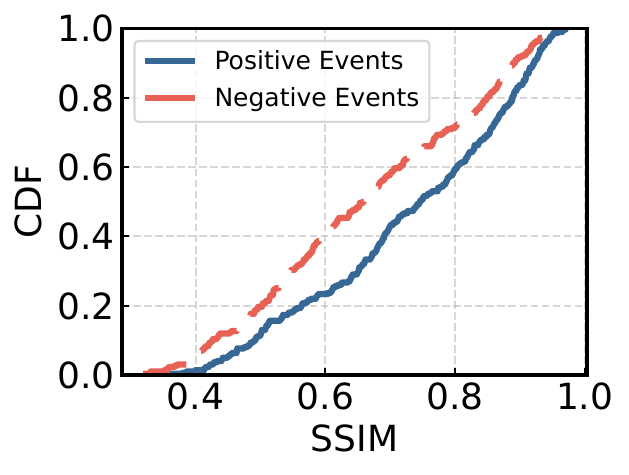}%
        \label{fig:ssim}%
    }
    \hfill
    \subfloat[CDF of NCC values.]{%
        \includegraphics[width=0.49\linewidth]{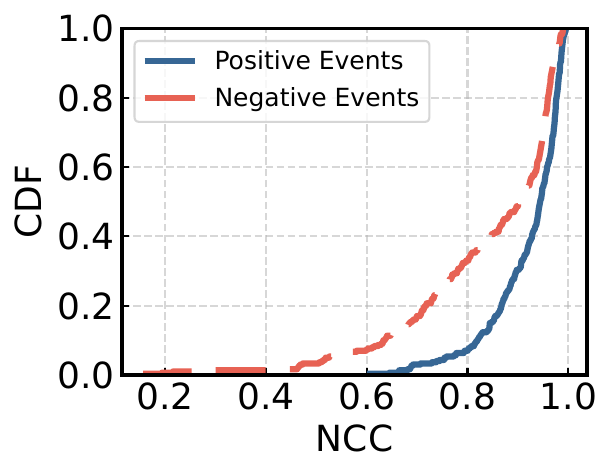}%
        \label{fig:ncc}%
    }
    \caption{CDF of similarity measurements.}
    \vspace{-0.2in}
    \label{fig:simcdf}
\end{figure}

\begin{figure}[htb]
    \centering
    \begin{minipage}[b]{0.48\linewidth}
        \includegraphics[height=0.8\linewidth]{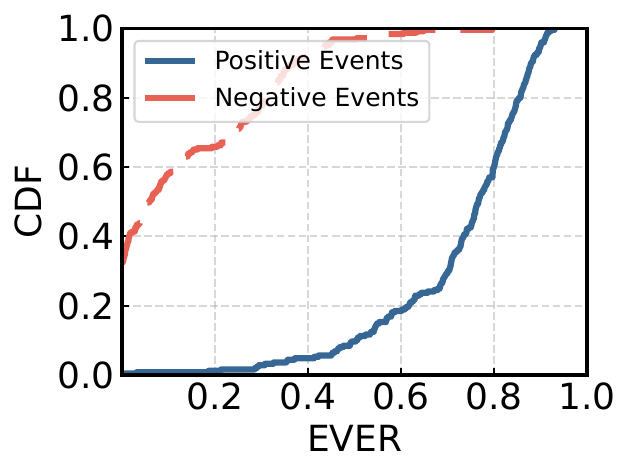}
	   \caption{CDF of IoU metrics.}
	   \label{fig:evercdf}
    \end{minipage}
    \hfill
    \begin{minipage}[b]{0.48\linewidth}
	   \includegraphics[height=0.8\linewidth]{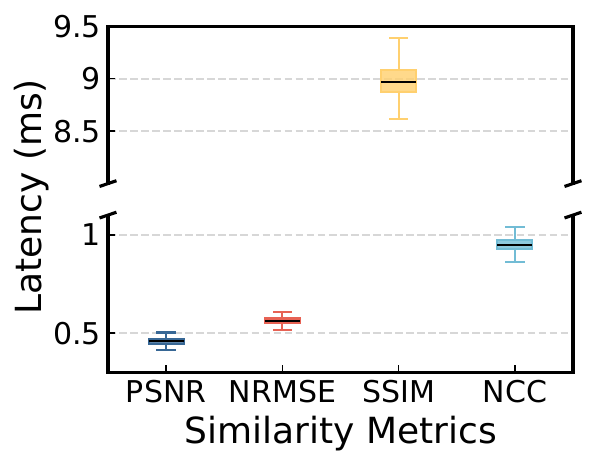}
	   \caption{Latency comparison.}
	   \label{fig:lat_sim_metrics}
    \end{minipage}
\end{figure}

Besides PSNR, there are other traditional metrics to measure the similarity between two images, like SSIM, NRMSE, and NCC. 
To explore the impact of different similarity metrics on the Sim-Ver baseline, we calculate those metrics on the custom dataset and draw their CDFs in \Cref{fig:simcdf}. We can observe that although all those metrics showcase obviously different CDFs between positive events and negative events, it will be difficult to differentiate them with a simple threshold-based policy. In other words, if we draw a vertical line on those figures, there will always be large overlapping areas of the positive and negative events regardless of the values. 
Specifically, the verification accuracy for those four metrics are 29\%, 65\%, 63.667\%, and 63.667\%, respectively. Note that the extremely poor performance of NRMSE indicates a biased distribution of NRMSE on the validation and test set.
In comparison, we can observe a clear divergence between IoU metrics for positive and negative events used in \sysname, as shown in \Cref{fig:evercdf}.
Although the verification accuracy for approaches using similarity metrics is undesirable, the calculation latencies for most similarity metrics are much lower than using ML models, as shown in \Cref{fig:lat_sim_metrics}. As such, those metrics may be useful in particular scenarios where the users are more latency-sensitive. Based on the accuracy and latency results, we utilize the PSNR metric in our baseline Sim-Ver. 

\vspace{-0.1in}

\section{Discussion} \label{sec::discuss}
In this section, we discuss several limitations of our proposed system and indicate possible directions for future work. 

\textbf{Generality.} The current framework is specifically designed and optimized for rigid objects. Extending the system to support non-rigid objects remains an open challenge and a direction for future research, as non-rigid objects exhibit variable shapes and deformations that complicate segmentation and verification.
On the other hand, our custom dataset and post-processing enable reliable results in controlled settings, but broader generalization would benefit from larger, more diverse datasets. For new tasks, the system currently requires task-specific reference images, which introduces setup overhead. Although this is a one-time effort common in practical workflows, future extensions may reduce this cost by leveraging large-scale datasets (e.g., LEGOGPT~\cite{pun2025generating} for LEGO tasks). Besides, integrating large language models (e.g., ContextDET~\cite{zang2025contextual}) could reduce setup effort by generating task steps, guiding reference frame creation and assisting target frame segmentation through multimodal understanding.
We consider these directions important for validating and expanding the system’s applicability to a wider range of MR applications.

\textbf{Robustness.} Our system exhibits robustness to moderate misalignment and lighting variations. Alignment is achieved through tag-based localization, which remains accurate as long as the tag is visible. However, in extreme scenarios, such as occlusion of both the tag and object, or when the viewpoint deviates significantly from the reference frame, alignment may fail due to undetectable tags. These limitations are inherent to vision-based systems that rely on visual markers. Potential enhancements include using multiple tags or placing tags on vertical surfaces to increase visibility across a wider range of viewing angles. For lighting variation, the segmentation model is trained with data covering diverse illumination conditions, enabling consistent performance in typical indoor environments. While the system performs reliably in standard settings, improving robustness under severe occlusion and extreme viewpoints remains a direction for future work.

\textbf{Partial Completion.} Our system exhibits a degree of capability to handle partial completion with manual control. Since the framework renders and verifies one assembly step at a time, users can manually navigate to a specific step using a control button. This functionality enables use cases such as resuming a task after interruption, continuing work initiated by another user, or recovering from an unexpected system reset. While this manual approach is functional, it may be less user-friendly in practice. A more seamless solution would involve automatically detecting the current completion state and inferring the correct step, which presents an important direction for future development.

\section{Conclusion}\label{sec::conclusion}
In this paper, we proposed \sysname: an edge-assisted auto-verification system for MR-aided operation applications. \sysname automatically captures reference and target frames based on the user's operation and uploads them to an edge server, where automatic verification is performed by segmenting virtual and physical objects using virtual information and the YOLO model, respectively.  
\sysname innovatively integrates the position of virtual objects to facilitate the verification process. Besides, \sysname employs a series of optimization techniques to ensure practicality, efficiency, and responsiveness. To evaluate its generality and accuracy, we generated two synthetic datasets based on public datasets and collected custom datasets, where \sysname achieves over 90\% auto-verification accuracy on all test sets. Moreover, evaluations of the system implementation demonstrated that \sysname achieves less than 100 ms end-to-end latency, low energy consumption, and low computational overhead compared to that without the auto-verification process. 

\section*{Acknowledgment}
This work has been supported in part by NSF under the grants CNS-2152658 and M3X-2420351, and DARPA grant HR0011-2420366.

\bibliographystyle{abbrv-doi}
\bibliography{refs}

\begin{thebibliography}{10}

\bibitem{abbas2017mobile}
N.~Abbas, Y.~Zhang, A.~Taherkordi, and T.~Skeie.
\newblock Mobile edge computing: A survey.
\newblock {\em IEEE Internet of Things Journal}, 5(1):450--465, 2017. doi: {{%
10\hspace{.1pt}\discretionary{.}{%
}{.}\hspace{.4pt}1109\discretionary{/}{%
}{/}JIOT\hspace{.1pt}\discretionary{.}{%
}{.}\hspace{.4pt}2017\hspace{.1pt}\discretionary{.}{%
}{.}\hspace{.4pt}2750180}}


\bibitem{alshowaish2022trademark}
H.~Alshowaish, Y.~Al-Ohali, and A.~Al-Nafjan.
\newblock Trademark image similarity detection using convolutional neural network.
\newblock {\em Applied Sciences}, 12(3):1752, 2022. doi: {{%
10\hspace{.1pt}\discretionary{.}{%
}{.}\hspace{.4pt}3390\discretionary{/}{%
}{/}app12031752}}


\bibitem{alur1991techniques}
R.~Alur.
\newblock {\em Techniques for automatic verification of real-time systems}.
\newblock stanford university, 1991.

\bibitem{brown1992survey}
L.~G. Brown.
\newblock A survey of image registration techniques.
\newblock {\em ACM computing surveys (CSUR)}, 24(4):325--376, 1992. doi: {{%
10\hspace{.1pt}\discretionary{.}{%
}{.}\hspace{.4pt}1145\discretionary{/}{%
}{/}146370\hspace{.1pt}\discretionary{.}{%
}{.}\hspace{.4pt}146374}}


\bibitem{chen2022enhancing}
J.~Chen, F.~Qian, and B.~Li.
\newblock Enhancing quality of experience for collaborative virtual reality with commodity mobile devices.
\newblock In {\em 2022 IEEE 42nd International Conference on Distributed Computing Systems (ICDCS)}, pp. 1018--1028. IEEE, 2022. doi: {{%
10\hspace{.1pt}\discretionary{.}{%
}{.}\hspace{.4pt}1109\discretionary{/}{%
}{/}ICDCS54860\hspace{.1pt}\discretionary{.}{%
}{.}\hspace{.4pt}2022\hspace{.1pt}\discretionary{.}{%
}{.}\hspace{.4pt}00102}}


\bibitem{chen2023motion}
J.~Chen, X.~Qin, G.~Zhu, B.~Ji, and B.~Li.
\newblock Motion-prediction-based wireless scheduling for interactive panoramic scene delivery.
\newblock {\em IEEE Transactions on Network Science and Engineering}, 11(2):1566--1579, 2023.

\bibitem{chen2015glimpse}
T.~Y.-H. Chen, L.~Ravindranath, S.~Deng, P.~Bahl, and H.~Balakrishnan.
\newblock Glimpse: Continuous, real-time object recognition on mobile devices.
\newblock In {\em Proceedings of the 13th ACM Conference on Embedded Networked Sensor Systems}, pp. 155--168, 2015. doi: {{%
10\hspace{.1pt}\discretionary{.}{%
}{.}\hspace{.4pt}1145\discretionary{/}{%
}{/}2809695\hspace{.1pt}\discretionary{.}{%
}{.}\hspace{.4pt}2809711}}


\bibitem{clarke1986automatic}
E.~M. Clarke, E.~A. Emerson, and A.~P. Sistla.
\newblock Automatic verification of finite-state concurrent systems using temporal logic specifications.
\newblock {\em ACM Transactions on Programming Languages and Systems (TOPLAS)}, 8(2):244--263, 1986. doi: {{%
10\hspace{.1pt}\discretionary{.}{%
}{.}\hspace{.4pt}1145\discretionary{/}{%
}{/}5397\hspace{.1pt}\discretionary{.}{%
}{.}\hspace{.4pt}5399}}


\bibitem{dong2023collaborative}
Z.~Dong, J.~Chen, and B.~Li.
\newblock Collaborative mixed-reality-based firefighter training.
\newblock In {\em IEEE INFOCOM 2023-IEEE Conference on Computer Communications Workshops (INFOCOM WKSHPS)}, pp. 1--2. IEEE, 2023. doi: {{%
10\hspace{.1pt}\discretionary{.}{%
}{.}\hspace{.4pt}1109\discretionary{/}{%
}{/}INFOCOMWKSHPS57453\hspace{.1pt}\discretionary{.}{%
}{.}\hspace{.4pt}2023\hspace{.1pt}\discretionary{.}{%
}{.}\hspace{.4pt}10226009}}


\bibitem{druzhkov2016survey}
P.~Druzhkov and V.~Kustikova.
\newblock A survey of deep learning methods and software tools for image classification and object detection.
\newblock {\em Pattern Recognition and Image Analysis}, 26(1):9--15, 2016. doi: {{%
10\hspace{.1pt}\discretionary{.}{%
}{.}\hspace{.4pt}1134\discretionary{/}{%
}{/}S1054661816010065}}


\bibitem{dumitru2023using}
R.-G. Dumitru, D.~Peteleaza, and C.~Craciun.
\newblock Using duck-net for polyp image segmentation.
\newblock {\em Scientific reports}, 13(1):9803, 2023. doi: {{%
10\hspace{.1pt}\discretionary{.}{%
}{.}\hspace{.4pt}1038\discretionary{/}{%
}{/}s41598\discretionary{%
}{-}{-}023\discretionary{%
}{-}{-}36940\discretionary{%
}{-}{-}5}}


\bibitem{dwyer2022roboflow}
B.~Dwyer, J.~Nelson, J.~Solawetz, et~al.
\newblock Roboflow (version 1.0)[software], 2022.

\bibitem{greengard2019virtual}
S.~Greengard.
\newblock {\em Virtual reality}.
\newblock Mit Press, 2019.

\bibitem{humanreaction}
Human {B}enchmark - {R}eaction {T}ime {T}est.

\bibitem{igarashi1975automatic}
S.~Igarashi, R.~L. London, and D.~C. Luckham.
\newblock Automatic program verification i: A logical basis and its implementation.
\newblock {\em Acta Informatica}, 4:145--182, 1975. doi: {{%
10\hspace{.1pt}\discretionary{.}{%
}{.}\hspace{.4pt}1007\discretionary{/}{%
}{/}BF00288746}}


\bibitem{Jocher_YOLO_by_Ultralytics_2023}
G.~Jocher, A.~Chaurasia, and J.~Qiu.
\newblock {YOLO by Ultralytics}, Jan. 2023.

\bibitem{kirillov2023segany}
A.~Kirillov, E.~Mintun, N.~Ravi, H.~Mao, C.~Rolland, L.~Gustafson, T.~Xiao, S.~Whitehead, A.~C. Berg, W.-Y. Lo, P.~Doll{\'a}r, and R.~Girshick.
\newblock Segment anything.
\newblock {\em arXiv:2304.02643}, 2023. doi: {{%
10\hspace{.1pt}\discretionary{.}{%
}{.}\hspace{.4pt}48550\discretionary{/}{%
}{/}arXiv\hspace{.1pt}\discretionary{.}{%
}{.}\hspace{.4pt}2304\hspace{.1pt}\discretionary{.}{%
}{.}\hspace{.4pt}02643}}


\bibitem{legoimporter}
Lego model importer.

\bibitem{liu2019edge}
L.~Liu, H.~Li, and M.~Gruteser.
\newblock Edge assisted real-time object detection for mobile augmented reality.
\newblock In {\em The 25th annual international conference on mobile computing and networking}, pp. 1--16, 2019. doi: {{%
10\hspace{.1pt}\discretionary{.}{%
}{.}\hspace{.4pt}1145\discretionary{/}{%
}{/}3300061\hspace{.1pt}\discretionary{.}{%
}{.}\hspace{.4pt}3300116}}


\bibitem{lu2007survey}
D.~Lu and Q.~Weng.
\newblock A survey of image classification methods and techniques for improving classification performance.
\newblock {\em International journal of Remote sensing}, 28(5):823--870, 2007. doi: {{%
10\hspace{.1pt}\discretionary{.}{%
}{.}\hspace{.4pt}1080\discretionary{/}{%
}{/}01431160600746456}}


\bibitem{makwana2023pcbsegclassnet}
D.~Makwana, S.~Mittal, et~al.
\newblock Pcbsegclassnet—a light-weight network for segmentation and classification of pcb component.
\newblock {\em Expert Systems with Applications}, 225:120029, 2023. doi: {{%
10\hspace{.1pt}\discretionary{.}{%
}{.}\hspace{.4pt}1016\discretionary{/}{%
}{/}j\hspace{.1pt}\discretionary{.}{%
}{.}\hspace{.4pt}eswa\hspace{.1pt}\discretionary{.}{%
}{.}\hspace{.4pt}2023\hspace{.1pt}\discretionary{.}{%
}{.}\hspace{.4pt}120029}}


\bibitem{mediacodec}
Android {M}edia{C}odec {API}.

\bibitem{naticchia2018mixed}
B.~Naticchia, A.~Corneli, A.~Carbonari, A.~Bonci, and M.~Pirani.
\newblock Mixed reality approach for the management of building maintenance and operation.
\newblock In {\em ISARC. Proceedings of the International Symposium on Automation and Robotics in Construction}, vol.~35, pp. 1--8. IAARC Publications, 2018. doi: {{%
10\hspace{.1pt}\discretionary{.}{%
}{.}\hspace{.4pt}22260\discretionary{/}{%
}{/}ISARC2018\discretionary{/}{%
}{/}0028}}


\bibitem{olson2011apriltag}
E.~Olson.
\newblock Apriltag: A robust and flexible visual fiducial system.
\newblock In {\em 2011 IEEE International Conference on Robotics and Automation}, pp. 3400--3407. IEEE, 2011. doi: {{%
10\hspace{.1pt}\discretionary{.}{%
}{.}\hspace{.4pt}1109\discretionary{/}{%
}{/}ICRA\hspace{.1pt}\discretionary{.}{%
}{.}\hspace{.4pt}2011\hspace{.1pt}\discretionary{.}{%
}{.}\hspace{.4pt}5979561}}


\bibitem{pun2025generating}
A.~Pun, K.~Deng, R.~Liu, D.~Ramanan, C.~Liu, and J.-Y. Zhu.
\newblock Generating physically stable and buildable lego designs from text.
\newblock {\em arXiv preprint arXiv:2505.05469}, 2025. doi: {{%
10\hspace{.1pt}\discretionary{.}{%
}{.}\hspace{.4pt}48550\discretionary{/}{%
}{/}arXiv\hspace{.1pt}\discretionary{.}{%
}{.}\hspace{.4pt}2505\hspace{.1pt}\discretionary{.}{%
}{.}\hspace{.4pt}05469}}


\bibitem{sahija2021impact}
D.~Sahija.
\newblock Impact of iot integration with mixed reality on manufacturing operations.
\newblock {\em Int. J. Res. Cult. Soc}, pp. 78--85, 2021.

\bibitem{shaik2015comparative}
K.~B. Shaik, P.~Ganesan, V.~Kalist, B.~Sathish, and J.~M.~M. Jenitha.
\newblock Comparative study of skin color detection and segmentation in hsv and ycbcr color space.
\newblock {\em Procedia Computer Science}, 57:41--48, 2015. doi: {{%
10\hspace{.1pt}\discretionary{.}{%
}{.}\hspace{.4pt}1016\discretionary{/}{%
}{/}j\hspace{.1pt}\discretionary{.}{%
}{.}\hspace{.4pt}procs\hspace{.1pt}\discretionary{.}{%
}{.}\hspace{.4pt}2015\hspace{.1pt}\discretionary{.}{%
}{.}\hspace{.4pt}07\hspace{.1pt}\discretionary{.}{%
}{.}\hspace{.4pt}362}}


\bibitem{simvgg16}
Image {S}imilarity {C}omparison using {VGG}16 {D}eep {L}earning {M}odel.

\bibitem{smith2016augmented}
M.~Smith, A.~Maiti, A.~D. Maxwell, and A.~A. Kist.
\newblock Augmented and mixed reality features and tools for remote laboratory experiment.
\newblock {\em International Journal of Online Engineering}, 12(7):45--52, 2016. doi: {{%
10\hspace{.1pt}\discretionary{.}{%
}{.}\hspace{.4pt}3991\discretionary{/}{%
}{/}ijoe\hspace{.1pt}\discretionary{.}{%
}{.}\hspace{.4pt}v12i07\hspace{.1pt}\discretionary{.}{%
}{.}\hspace{.4pt}5851}}


\bibitem{thoravi2019loki}
B.~Thoravi~Kumaravel, F.~Anderson, G.~Fitzmaurice, B.~Hartmann, and T.~Grossman.
\newblock Loki: Facilitating remote instruction of physical tasks using bi-directional mixed-reality telepresence.
\newblock In {\em Proceedings of the 32nd Annual ACM Symposium on User Interface Software and Technology}, pp. 161--174, 2019. doi: {{%
10\hspace{.1pt}\discretionary{.}{%
}{.}\hspace{.4pt}1145\discretionary{/}{%
}{/}3332165\hspace{.1pt}\discretionary{.}{%
}{.}\hspace{.4pt}3347872}}


\bibitem{thung2009survey}
K.-H. Thung and P.~Raveendran.
\newblock A survey of image quality measures.
\newblock In {\em 2009 international conference for technical postgraduates (TECHPOS)}, pp. 1--4. IEEE, 2009. doi: {{%
10\hspace{.1pt}\discretionary{.}{%
}{.}\hspace{.4pt}1109\discretionary{/}{%
}{/}TECHPOS\hspace{.1pt}\discretionary{.}{%
}{.}\hspace{.4pt}2009\hspace{.1pt}\discretionary{.}{%
}{.}\hspace{.4pt}5412098}}


\bibitem{unityshader}
Unity - {M}anual: {W}riting shaders for different graphics {API}s.

\bibitem{van2019deep}
F.~Van~Beers, A.~Lindstr{\"o}m, E.~Okafor, and M.~Wiering.
\newblock Deep neural networks with intersection over union loss for binary image segmentation.
\newblock In {\em Proceedings of the 8th international conference on pattern recognition applications and methods}, pp. 438--445. SciTePress, 2019. doi: {{%
10\hspace{.1pt}\discretionary{.}{%
}{.}\hspace{.4pt}48550\discretionary{/}{%
}{/}10\hspace{.1pt}\discretionary{.}{%
}{.}\hspace{.4pt}5220\discretionary{/}{%
}{/}0007347504380445}}


\bibitem{videocodec}
{NVIDIA} {V}ideo {C}odec {SDK}.

\bibitem{wang2016apriltag}
J.~Wang and E.~Olson.
\newblock Apriltag 2: Efficient and robust fiducial detection.
\newblock In {\em 2016 IEEE/RSJ International Conference on Intelligent Robots and Systems (IROS)}, pp. 4193--4198. IEEE, 2016. doi: {{%
10\hspace{.1pt}\discretionary{.}{%
}{.}\hspace{.4pt}1109\discretionary{/}{%
}{/}IROS\hspace{.1pt}\discretionary{.}{%
}{.}\hspace{.4pt}2016\hspace{.1pt}\discretionary{.}{%
}{.}\hspace{.4pt}7759617}}


\bibitem{xie2021segformer}
E.~Xie, W.~Wang, Z.~Yu, A.~Anandkumar, J.~M. Alvarez, and P.~Luo.
\newblock Segformer: Simple and efficient design for semantic segmentation with transformers.
\newblock {\em Advances in neural information processing systems}, 34:12077--12090, 2021. doi: {{%
10\hspace{.1pt}\discretionary{.}{%
}{.}\hspace{.4pt}48550\discretionary{/}{%
}{/}arXiv\hspace{.1pt}\discretionary{.}{%
}{.}\hspace{.4pt}2105\hspace{.1pt}\discretionary{.}{%
}{.}\hspace{.4pt}15203}}


\bibitem{yan2021augmented}
W.~Yan.
\newblock Augmented reality instructions for construction toys enabled by accurate model registration and realistic object/hand occlusions.
\newblock {\em Virtual Reality}, pp. 1--14, 2021. doi: {{%
10\hspace{.1pt}\discretionary{.}{%
}{.}\hspace{.4pt}1007\discretionary{/}{%
}{/}s10055\discretionary{%
}{-}{-}021\discretionary{%
}{-}{-}00582\discretionary{%
}{-}{-}7}}


\bibitem{yao2023design}
X.~Yao.
\newblock The design and scalable implementation of collaborative mixed-reallity-aided lego creation.
\newblock Master's thesis, The Pennsylvania State University, 2023.

\bibitem{yao2022scalable}
X.~Yao, J.~Chen, T.~He, J.~Yang, and B.~Li.
\newblock A scalable mixed reality platform for remote collaborative lego design.
\newblock In {\em IEEE INFOCOM 2022-IEEE Conference on Computer Communications Workshops (INFOCOM WKSHPS)}, pp. 1--2. IEEE, 2022. doi: {{%
10\hspace{.1pt}\discretionary{.}{%
}{.}\hspace{.4pt}1109\discretionary{/}{%
}{/}INFOCOMWKSHPS54753\hspace{.1pt}\discretionary{.}{%
}{.}\hspace{.4pt}2022\hspace{.1pt}\discretionary{.}{%
}{.}\hspace{.4pt}9798010}}


\bibitem{zang2025contextual}
Y.~Zang, W.~Li, J.~Han, K.~Zhou, and C.~C. Loy.
\newblock Contextual object detection with multimodal large language models.
\newblock {\em International Journal of Computer Vision}, 133(2):825--843, 2025. doi: {{%
10\hspace{.1pt}\discretionary{.}{%
}{.}\hspace{.4pt}1007\discretionary{/}{%
}{/}s11263\discretionary{%
}{-}{-}024\discretionary{%
}{-}{-}02214\discretionary{%
}{-}{-}4}}


\end{thebibliography}
\end{document}